\documentclass[preprint2]{aastex}

\usepackage{apjfonts}
\usepackage{mathptmx}

\slugcomment{Accepted for publication in {\it The Astronomical Journal}}

\shorttitle{DEBRIS DISK AROUND HD~92945}
\shortauthors{GOLIMOWSKI ET AL.}
 
\begin{document}
 
\title{Hubble and Spitzer Space Telescope Observations of \\ the Debris Disk around the Nearby K Dwarf HD 92945\altaffilmark{1}}

\author{
D.~A.~Golimowski\altaffilmark{2},
J.~E.~Krist\altaffilmark{3},
K.~R.~Stapelfeldt\altaffilmark{3},
C.~H.~Chen\altaffilmark{2},
D.~R.~Ardila\altaffilmark{4},
G.~Bryden\altaffilmark{3},
M.~Clampin\altaffilmark{5},
H.~C.~Ford\altaffilmark{6},
G.~D.~Illingworth\altaffilmark{7},
P.~Plavchan\altaffilmark{8},
G.~H.~Rieke\altaffilmark{9},
and
K.~Y.~L.~Su\altaffilmark{9}
}

\altaffiltext{1}{ 
Based in part on guaranteed observing time awarded by the National Aeronautics and Space Administration (NASA) to the Advanced 
Camera for Surveys Investigation Definition Team and the Multiband Imaging Photometer for Spitzer Instrument Team.
}
\altaffiltext{2}{ 
Space Telescope Science Institute, 
3700 San Martin Drive, 
Baltimore, MD 21218
}
\altaffiltext{3}{ 
Jet Propulsion Laboratory,
California Insitute of Technology,
4800 Oak Grove Drive,
Mail Stop 183-900,
Pasadena, CA 91109
}
\altaffiltext{4}{ 
NASA Herschel Science Center,
California Institute of Technology,
Mail Stop 220-6,
Pasadena, CA 91125
}
\altaffiltext{5}{ 
NASA Goddard Space Flight Center,
Code 681, 
Greenbelt, MD 20771 
}
\altaffiltext{6}{ 
Department of Physics and Astronomy, 
The Johns Hopkins University, 
3400 North Charles Street, 
Baltimore, MD 21218
}
\altaffiltext{7}{ 
Lick Observatory,
University of California at Santa Cruz,
1156 High Street,
Santa Cruz, CA 95064
}
\altaffiltext{8}{ 
NASA Exoplanet Science Institute,
California Institute of Technology,
770 South Wilson Avenue, 
Pasadena, CA 91106
}
\altaffiltext{9}{
Steward Observatory,
University of Arizona,
933 North Cherry Avenue,
Tucson, AZ 85721
}

\begin{abstract} 
We present the first resolved images of the debris disk around the nearby K dwarf HD~92945, obtained with the {\it Hubble Space 
Telescope's} Advanced Camera for Surveys.  Our F606W (Broad $V$) and F814W (Broad~$I$) coronagraphic images reveal an inclined,
axisymmetric disk consisting of an inner ring about 2\farcs0--3\farcs0 \linebreak (43--65~AU) from the star and an extended outer disk whose
surface brightness declines slowly with increasing radius approximately 3\farcs0--5\farcs1 (65--110~AU) from the star.  A precipitous
drop in the surface brightness beyond 110~AU suggests that the outer disk is truncated at that distance.  The radial surface-density 
profile is peaked at both the inner ring and the outer edge of the disk.  The dust in the outer disk scatters neutrally but 
isotropically, and it has a low $V$-band albedo of 0.1.   This combination of axisymmetry, ringed and extended morphology, and 
isotropic neutral scattering is unique among the 16 debris disks currently resolved in scattered light.  We also present 
new infrared photometry and spectra of HD~92945 obtained with the {\it Spitzer Space Telescope's} Multiband Imaging Photometer and 
InfraRed Spectrograph.  These data reveal no infrared excess from the disk shortward of 30~\micron\ and constrain the width of the 
70~\micron\ source to $\lesssim 180$~AU.  Assuming that the dust comprises compact grains of astronomical silicate with a 
surface-density profile described by our scattered-light model of the disk, we successfully model the 24--350~\micron\ emission with 
a minimum grain size of $a_{min} = 4.5$~\micron\ and a size distribution proportional to $a^{-3.7}$ throughout the disk, but 
with maximum grain sizes of 900~\micron\ in the inner ring and 50~\micron\ in the outer disk.  Together, our {\it HST}
and {\it Spitzer} observations indicate a total dust mass of $\sim 0.001~M_{\earth}$.  However, our observations provide contradictory 
evidence of the dust's physical characteristics:  its neutral $V$--$I$ color and lack of 24~\micron\ emission imply grains larger 
than a few microns, but its isotropic scattering and low albedo suggest a large population of submicron-sized grains.  If grains 
smaller than a few microns are absent, then stellar radiation pressure may be the cause only if the dust is composed of highly 
absorptive materials like graphite.  The dynamical causes of the sharply edged inner ring and outer disk are unclear, but recent 
models of dust creation and transport in the presence of migrating planets support the notion that the disk indicates an advanced 
state of planet formation around HD~92945.
\end{abstract}

 
\section{Introduction\label{intro}}

A circumstellar disk is called a ``debris disk'' if the age of its host star exceeds the lifetimes of its constituent dust 
grains.  Intergrain collisions, radiation pressure, and Poynting--Robertson drag eliminate dust around solar-type stars in 
$\sim 10$~Myr \citep{bac93,lag00}; \linebreak \citep{zuc01,mey07,wya08}.  The presence of dust around older stars suggests that the dust has 
been replenished by cometary evaporation or by collisions among unseen planetesimals.  Resolved images of debris disks enable 
studies of the properties and dynamics of grains that compose the debris in extrasolar planetary systems.  They also provide 
opportunities for indirectly detecting planets via their dynamical effects on the distribution of the dust.

Observing light scattered by debris disks is difficult because the disks have small optical depths ($\tau \lesssim 10^{-3}$) 
and their host stars are bright.  Dozens of optically thick disks around young stars have been imaged in scattered light, but
only 16 debris disks have been resolved in this manner (Table~\ref{16disks}).  Five of these debris disks surround 
stars with masses $\lesssim 1~M_{\sun}$: AU~Microscopii \citep[spectral type M1~Ve;][]{kal04}, HD~107146 \citep[G2~V;][]{ard04},
HD~53143 \citep[K1~V;][]{kal06}, HD~61005 \citep[G8~V;][]{hin07}, and HD~92945 (K1~V; this paper).  Thus, few examples
of the circumstellar environments of low-mass stars in the late stages of planet formation are known.

The nearby star HD~92945 (K1~V, $V =  7.72$, distance = $21.57 \pm 0.39$~pc; \citealt{esa97}) was first identified by \citet{sil00}
as a candidate for attendant circumstellar dust based on its 60~\micron\ flux measured by the {\it Infrared Astronomical Satellite 
(IRAS)}, which significantly exceeds the expected photospheric flux at that wavelength.  Mid-infrared and submillimeter images of 
HD~92945 from the {\it Spitzer Space Telescope} and the Caltech Submillimeter Observatory also reveal excess fluxes at 70~\micron\
and 350~\micron\ that, along with the {\it IRAS} measurement, are consistent with an optically thin disk of blackbody grains having 
an inner radius of 15--23~AU, an equilibrium temperature of 40--46~K, and a fractional infrared luminosity of $L_{\it IR}/L_{\rm *} 
= 7.7 \times 10^{-4}$, where $L_{\rm *}$ is the bolometric luminosity of the star \citep{che05,pla09}.  This fractional luminosity 
is about half that of the disk surrounding the $\sim 12$~Myr-old, A5~{\small V} star $\beta$~Pictoris \citep{dec03}.  \citet{lop06} 
estimate an age of $\sim 80$--150~Myr for HD~92945 based on its luminosity, lithium absorption strength, and association with the
Local (Pleiades) Moving Group.  However, \citet{pla09} suggest an age of $\sim 300$~Myr based on correlations of age with X-ray 
and Ca~{\small II} H and K emission \citep{mam08}.

In this paper, we present the first resolved scattered-light images of the debris disk around HD~92945.  These coronagraphic
images from the {\it Hubble Space Telescope (HST)} provide only the second opportunity to study a debris disk around a K dwarf
at visible wavelengths.  We also present new mid-infrared photometry and spectra of the disk obtained from the {\it Spitzer 
Space Telescope}.  Together, these {\it HST} and {\it Spitzer} observations allow us to constrain the optical properties, sizes,
and spatial distribution of the dust grains within the disk.  We use these constraints to revise the estimates of the disk's 
fractional infrared luminosity and mass previously reported by \citet{che05} and \citet{pla09}.   Finally, we discuss the 
morphology of the disk in the context of other spatially resolved debris disks and current models of dust production and dynamics 
during the late stages of planet formation.

\section{Observations and Data Processing\label{observations}}

Our {\it HST} and {\it Spitzer} observations are presented jointly as the result of agreements between the {\it HST} Advanced Camera 
for Surveys (ACS) Investigation Definition Team (IDT), the Multiband Imaging Photometer for {\it Spitzer} (MIPS) Instrument Team, 
and the investigators associated with {\it Spitzer} Fellowship program 241 (C.~H.~Chen, Principal Investigator), which utilizes 
{\it Spitzer's} InfraRed Spectrograph (IRS) and the MIPS Spectral Energy Distribution (SED) mode.  The inclusion of HD~92945 in the 
ACS IDT's dedicated effort to discover and characterize nearby debris disks was informed by an early 70~\micron\ MIPS detection of 
cold circumstellar dust around the star \citep{che05}.  The MIPS detection confirmed the previous {\it IRAS} measurement and established
an acceptible likelihood of detection by ACS based on previous successful imaging of the debris disk around HD~107146 \citep{ard04}.

\subsection{HST/ACS Coronagraphic Observations\label{acsobs}}

\subsubsection{Imaging Strategy and Reduction\label{acsred}}

The ACS observations of HD~92945 were performed as part of {\it HST} Guaranteed Time Observer (GTO) program 10330, which followed the 
IDT's standard strategy of first obtaining coronagraphic images through a single broadband filter to assess the presence of a debris 
disk and later obtaining multiband images of the disk, if warranted.  The first images of HD~92945 were obtained on UT~2004 December~1 
using the High Resolution Channel (HRC) of ACS \citep{for03,may10}.  The HRC has a $1024 \times 1024$-pixel CCD detector
whose pixels subtend an area of 0\farcs028~$\times$~0\farcs025, providing a $\sim 29'' \times 26''$ field of view (FOV).  HD~92945 was 
acquired in the standard ``peak-up'' mode with the coronagraph assembly deployed in the focal plane of the aberrated beam.  The star was
then positioned behind the small (0\farcs9 radius) occulting spot located approximately at the center of the FOV.  Two successive 
coronagraphic exposures of 1100~s were recorded using the F606W (Broad $V$) filter.  Two 0.1~s and two 35~s direct (i.e., non-coronagraphic)
exposures of HD~92945 were also recorded for the purposes of photometric calibration and detection of any bright circumstellar material 
obscured by the occulting spot.  The 0.1~s exposures were digitized using an analog-to-digital (A/D) gain of $4~e^-$~DN$^{-1}$ to allow 
unsaturated images of the unocculted star; the other images were recorded with an A/D gain of $2~e^-$~DN$^{-1}$ for better sampling of 
faint sources.  All images were recorded within one {\it HST} orbit.  A similar set of F606W images of HD~100623 (K0~V, $V = 5.96$, 
$d = 9.54 \pm 0.07$~pc; \citealt{esa97}) was recorded during the orbit immediately following the observation of HD~92945.  These images 
provide references for the instrumental point-spread functions (PSFs) of a nearby star with colors similar to HD~92945 but having no 
known circumstellar dust \citep{smi92}.

Follow-up HRC observations of HD~92945 were conducted on UT~2005 July~12.  (The same observations attempted previously on UT~2005 
April~21 were unsatisfactory because of failed guide-star acquisition.)  Because of changing seasonal constraints on the spacecraft's
roll angle, the FOV was rotated counterclockwise by 157\arcdeg\ from the orientation obtained 7 months earlier.  Multiple coronagraphic 
exposures totalling 4870~s and 7600~s were recorded through the F606W and F814W (Broad~$I$) filters, respectively, over 5 consecutive 
{\it HST} orbits.  One direct exposure of 0.1~s was also recorded through each filter for photometric calibration.  The A/D gain settings
conformed with those used in the first-epoch observation.  Likewise, coronagraphic exposures of HD~100623 totalling 2178~s and 2412~s 
were recorded through F606W and F814W, respectively, over 2 consecutive orbits immediately following the observation of HD~92945.  One 
10~s coronagraphic exposure of HD~100623 was recorded through each filter to ensure unsaturated imaging of the PSF along the perimeter 
of the occulting spot.  One direct exposure of 0.1~s was also recorded through each filter for photometric purposes.

The initial stages of image reduction (i.e., subtraction of bias and dark frames and division by a direct or coronagraphic flat 
field) were performed by the ACS image calibration pipeline at the Space Telescope Science Institute (STScI; \citealt{pav06}). 
We averaged the images recorded with the same combinations of filter, exposure time, and roll angle after interpolating over 
permanent bad pixels and rejecting transient artifacts identified as statistical outliers.  We then normalized the averaged images 
to unit exposure time.  We replaced saturated pixels in the long-exposure images of HD~100623 with unsaturated pixels at 
corresponding locations in the 10~s images.  Throughout this process, we tracked the uncertainties associated with each image 
pixel.  In this manner, we created cosmetically clean, high-contrast images and meaningful error maps for each combination of 
star, filter, and roll angle.  The reduced, second-epoch F606W images of HD~92945 and HD~100623 are shown in the top and middle
panels of Figure~\ref{f606w}, respectively.  The corresponding first-epoch F606W and second-epoch F814W images are qualitatively
similar to those shown in Figure~\ref{f606w}.

\subsubsection{Subtraction of the Coronagraphic PSF\label{acspsfsub}}

\begin{figure*}[t]
  \epsscale{0.75}\plotone{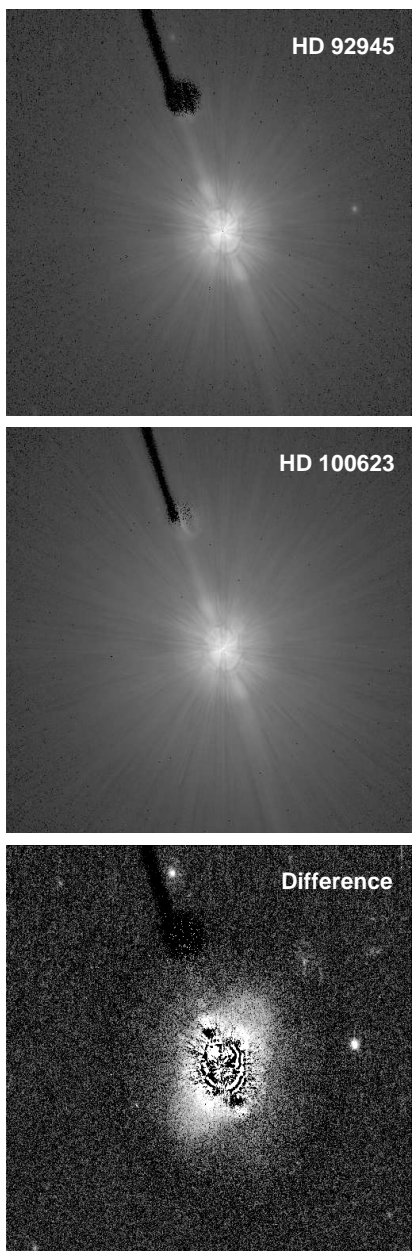}
   \caption{
	Reduced F606W (Broad $V$) images of HD~92945 {\it (top)} and the PSF-reference star HD~100623 {\it (middle)} obtained 
	with the ACS HRC coronagraph on UT~2005 July~12.  The linear scattered-light feature extending from the tip of the 
	occulting bar is an intrinsic component of the coronagraphic PSF.  The bottom panel shows the image of HD~92945 after 
	normalization, registration, and subtraction of the reference PSF.  This ``difference'' image reveals HD~92945's dusty
	debris disk, which is partly obscured by PSF-subtraction residuals surrounding the central occulting spot.  The 
	overlapping shadows of the HRC's occulting bar and large occulting spot are seen protruding from the top of each image.
	All images are displayed with logarithmic scaling and $2 \times 2$-pixel binning, but without correction of geometric 
	distortion.  
	}
  \label{f606w}
\end{figure*}

To examine the scattered-light component of the dust around HD~92945, we removed from each image the overwhelming light from the
occulted star's PSF.  By observing HD~92945 and HD~100623 in consecutive {\it HST} orbits, we limited the differences between the 
PSFs of the two stars that would otherwise be caused by inconsistent deployment of the coronagraph assembly, gradual migration of 
the occulting spot, or changes in {\it HST's} thermally driven focus cycles \citep[commonly called ``breathing'';][]{kri02}.  We 
measured the positions of the stars behind the occulting spot using the central peaks of the reduced coronagraphic PSFs 
(Figure~\ref{f606w}) that resulted from the reimaging of incompletely occulted, spherically aberrated starlight by ACS's corrective 
optics \citep{kri00}.  The positions of the two stars differed by $\sim 0.3$--0.5~pixel ($\sim 0$\farcs008--0\farcs013).  Similar 
offsets between two coronagraphic images of the same star cause $\sim 1$\% variations in the surface brightness profiles of the 
occulted PSFs over most of the HRC's FOV \citep{kri00}.  For different stars, this alignment error may be compounded by differences 
between the stars' colors and brightnesses.

Optimal subtraction of the coronagraphic PSF requires accurate normalization and registration of the filter images of the 
reference star HD~100623 with the corresponding images of HD~92945.  We used the 0.1~s direct F606W and F814W images and 
conventional aperture photometry to measure the relative brightnesses of each star in each bandpass.  The first-epoch F606W
images of HD~100623 were slightly saturated, but the flux-conserving A/D gain of 4$~e^-$~DN$^{-1}$ \citep{gil04} and a large 
(0\farcs5 radius) aperture permitted accurate measurement of the integrated flux.  We computed flux ratios, 
$F_{\rm HD100623}/F_{\rm HD92945}$, of 5.00 and 4.83 for F606W and F814W, respectively.  We divided the images of HD~100623 by 
these ratios to bring the integrated brightnesses of the reference PSFs into conformity with those of HD~92945. 

We aligned the normalized images of HD~100623 with the corresponding images of HD~92945 using an interactive routine that 
permits orthogonal shifts of an image with subpixel resolution and cubic convolution interpolation.  The shift intervals and 
normalization factors (i.e., $F_{\rm HD100623}/F_{\rm HD92945}$) were progressively refined throughout the iterative process.  We 
assessed the quality of the normalization and registration by visually inspecting the difference image created after each shift 
or normalization adjustment.  Convergence was reached when the subtraction residuals were visibly minimized and refinements of
the shift interval or normalization factor had inconsequential effects (e.g., bottom panel of Figure~\ref{f606w}).  Based on 
these qualitative assessments, we estimate that the uncertainty of the registration along each axis is $\pm 0.0625$~pixel and 
the uncertainty of $F_{\rm HD100623}/F_{\rm HD92945}$ in each bandpass is $\pm 1$\%. 

After subtracting the coronagraphic PSFs from each image of HD~92945, we transformed the images to correct the pronounced geometric distortion
across the HRC's image plane.  In doing so, we used the coefficients of the biquartic-polynomial distortion map provided by STScI
\citep{meu02} and cubic convolution interpolation to conserve the imaged flux.  This correction yields a rectified pixel scale of 
0\farcs025~$\times$~0\farcs025.  We then combined the first- and second-epoch F606W images by rotating the former images 
counterclockwise by 157\arcdeg, aligning the images according to the previously measured stellar centroids, and averaging the 
images after rejecting pixels that exceeded their local $3\sigma$ values.  Regions shaded by the HRC's occulting bar and large 
occulting spot were excluded from the average, as were regions corrupted by incomplete subtraction of the linear PSF artifact seen 
in Figure~\ref{f606w}.  (Because the orientations of the first- and second-epoch F606W images differed by 157\arcdeg, the 
regions obscured by the occulters and the linear artifact at one epoch were unobscured at the other epoch.)  Again, we tracked 
the uncertainties associated with each stage of image processing to maintain a meaningful map of random pixel errors.  

We combined in quadrature the final random-error maps with estimates of the systematic errors caused by uncertainties in the 
normalization and registration of the reference PSFs.  The systematic-error maps represent the convolved differences between the 
optimal PSF-subtracted image of HD~92945 and three nonoptimal ones generated by purposefully misaligning (along each axis) or 
misscaling the images of HD~100623 by amounts equal to our estimated uncertainties in PSF registration and 
$F_{\rm HD100623}/F_{\rm HD92945}$.  The total systematic errors are 1--5 times larger than the random errors within $\sim 1$\farcs5
of HD~92945, but they diminish to 2--25\% of the random errors beyond $\sim 3''$--$4''$ from the star.  We refer to the combined 
maps of random and systematic errors as total-error maps.  Note that these maps exclude systematic errors caused by changes in 
{\it HST's} focus between the observations of HD~92945 and HD~100623 or by small differences in the stars' colors, which affect 
the spatial and spectral distributions of light within the PSFs. \citep{kri00,kri03}.  Unfortunately, without precise knowledge 
of {\it HST's} thermally varying optical configuration or the stars' spectra, these systematic errors cannot be accurately 
predicted or assessed.

\begin{figure*}[t]
  \epsscale{2.20}\plotone{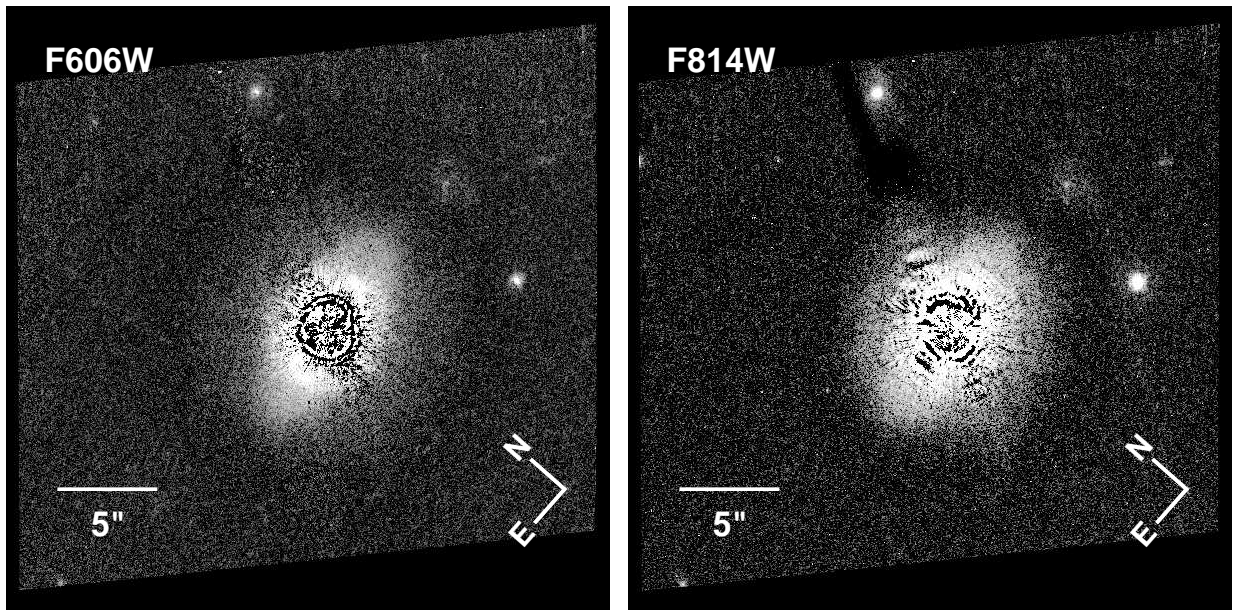}
   \caption{
	F606W and F814W images of HD~92945 after subtraction of the coronagraphic PSF and correction of HRC's geometric 
	distortion, but before removal of the wide-angle halo.  Both images are displayed with logarithmic scaling and 
	$2 \times 2$-pixel binning.  The orientation of the FOV reflects {\it HST's} roll angle during the second-epoch 
	observations.  The F606W image is the average of the images recorded at both observing epochs, excluding regions 
	obscured by the occulting bar, the large occulting spot, and subtraction residuals from the linear PSF artifact
	seen in Figure~\ref{f606w}.  No F814W images were recorded at the first epoch, so the regions affected by these
	artifacts remain.  The $5''$ scale bar corresponds to a projected distance of $108 \pm 2$~AU.
	}
  \label{psfsub}
\end{figure*}

Figure~\ref{psfsub} shows the PSF-subtracted and distortion-corrected images of HD~92945 in F606W and F814W.  Each image has been 
divided by the measured brightness of the star in the respective bandpass.  The alternating light and dark bands around the 
occulting spot reflect imperfect PSF subtraction caused by the slightly mismatched centroids of HD~92945 and HD~100623 
\citep{kri03}.  These subtraction residuals preclude accurate photometry of the disk within $\sim 2$\farcs0 ($\sim 43$~AU) of the 
star.  Likewise, the residual bands extending from the occulting bar in the F814W image are caused by imperfect subtraction of the
linear PSF artifact seen in Figure~\ref{f606w}.  (Similar residual bands were removed from the F606W image, as described above.)  
The images show three distinct structures centered about the star: (1) a relatively bright but partially obscured elliptical ring 
extending $\sim 2$\farcs0--3\farcs0 from the star, (2) an elliptical disk of lower but more uniform surface brightness extending 
beyond the ring and oriented at a position angle (PA) of $\sim 100$\arcdeg, and (3) a circular halo of very-low surface brightness 
encompassing both the ring and disk and extending outward to $\sim 6''$ and $7''$ in the F606W and F814W images, respectively.  
The halo itself is surrounded by a $\sim 2$\farcs5-wide ring having slightly lower surface brightness than the residual sky at the
edges of the FOV.

The faint extended halo and surrounding dark ring are well-known artifacts of the subtraction of two coronagraphic HRC images 
recorded at different phases of {HST's} breathing cycle \citep{kri00,kri03}.  The prominent halo in the F814W image may also
reflect incomplete subtraction of red photons ($\lambda \gtrsim 0.7$~\micron) that have diffusely scattered from the CCD substrate 
back into the CCD \citep{sir05}.  Some of this red halo may remain if the far-red colors of HD~92945 and HD~100623 differ slightly.  
To remove the halos from our F606W and F814W images, we azimuthally subtracted 9th- and 10th-order polynomials fitted to the average radial 
profiles of the halos measured beyond 3\farcs0 and 3\farcs2 from the star, respectively, and within $\pm 30$\arcdeg\ of the minor 
axis of the elliptical disk.  The accuracy of this subtraction is clearly contingent on the assumptions that the halo is 
azimuthally symmetric and that any actual circumstellar flux in the fitted region is negliglible.  The halo-subtracted images are 
shown in Figure~\ref{subhalo}; they represent the final stage of our HRC image processing.

\subsection{{\it Spitzer} Observations\label{spitzerobs}}

\subsubsection{MIPS Photometry\label{mipsphot}}

Following the 24~\micron\ and 70~\micron\ MIPS detections of HD~92945 recorded with the coarse (default) image resolution 
\citep{che05,pla09}, we obtained 70~\micron\ fine-scale (5\farcs24~$\times$~5\farcs33 per pixel) and 160~\micron\ coarse-scale 
(15\farcs96~$\times$~18\farcs04 per pixel) MIPS images as part of {\it Spitzer} GTO program 40679.  The 70~\micron\ fine-scale images 
were recorded on UT 2008~February~16 using two cycles of 10~s exposures and a small field dither pattern located at each of 4 target 
positions on a square grid with sides of 16\farcs22 (3.16~pixels).  The total effective on-source exposure time was 672~s.  The 
160~\micron\ images were obtained on UT 2008~June~22 using one cycle of the small field dither pattern and 10~s exposures at each 
of 9 target positions with relative offsets of $0''$ and $\pm 36''$ ($\pm 2.5$ pixels) along the short axis of the 2$\times$20 pixel 
array and with $0''$, $72''$, and $140''$ offsets along the long axis of the array to provide a large area for background measurement.
For both imaging bandpasses, the observing strategy provided enhanced subpixel sampling for improved PSF subtraction and deconvolution.

The 70~\micron\ and 160~\micron\ data were processed using the MIPS Data Analysis Tool (DAT; \citealt{gor05}) version 3.10, which
produces mosaics with rectified pixel sizes of 2\farcs62 and $8''$, respectively.  No field sources brighter than 3\% of HD~92945's
brightness were detected within $2'$ of the star.  Photometry within a 12-pixel square aperture yielded a 70~\micron\ flux density of 
$278 \pm 42$~mJy after subtraction of the local background, which is consistent with the default-scale measurement of \citet{pla09}.  
After applying a 7\% color correction relative to a fiducial Rayleigh--Jeans spectrum within the 70~\micron\ bandpass, we obtain a 
final 70~\micron\ flux density of $298 \pm 45$~mJy for HD~92945.  We derived a 160~\micron\ flux density of $285 \pm 34$~mJy using a 
12-pixel square aperture and an aperture correction of 0.73 derived from an Tiny Tim/Spitzer model PSF \citep{kri06}.  In both bandpasses,
the formal error is dominated by the absolute photometric calibration.  The near-infrared spectral leak of the MIPS 160~\micron\ bandpass
is not a concern for HD~92945, which has a relatively faint $K$-band magnitude of 5.7 \citep{col10}.

\subsubsection{MIPS and IRS Spectroscopy\label{mipsirsobs}}

We observed HD~92945 with the MIPS SED mode on UT 2005~June~23.  This mode produces a low-resolution ($R \approx 15$--25) spectrum 
from 55 to 90~\micron\ \citep{hei98,col10}.  The $120'' \times 20''$ slit was oriented at PA~=~300\arcdeg, i.e., $\sim 20$\arcdeg\ from 
the projected major axis of the disk seen in the ACS/HRC images.  Ten cycles of 10~s exposures were recorded for a total integration 
time of 629~s.  We obtained a background SED by chopping $1'$ from the nominal pointing, and we then subtracted the background 
SED from the target SED.  The images were reduced using MIPS DAT version 3.10, which produces a mosaic with 44~pixels 
$\times~4$\farcs9~pixel$^{-1}$ in the spatial direction and 65 pixels in the spectral direction \citep{gor05}.  We smoothed the 
mosaic in the spectral direction with a 5-pixel boxcar to improve the measured signal-to-noise ratio (S/N).  We then summed the 
signal in 5 columns centered on the peak emission along each row of the mosaic and extracted the SED.  To calibrate the flux,
we scaled a similarly reduced and extracted SED of the spectrophotometric standard star Canopus (spectral type G8~III) to a flux
density of 3.11~Jy at 70~\micron\ and then corrected for the MIPS spectral response function by assuming that Canopus has a 
Rayleigh-Jeans SED at far-infrared wavelengths.   We used a point-source aperture correction, keeping in mind that a secondary 
correction is needed when fitting a spatially extended disk model to the SED (\S\ref{irmodel}).

We performed IRS observations of HD~92945 on UT~2005 May~21 using both Short--Low (SL1: 7.4--14.5~\micron; SL2: 5.2--8.7~\micron) and 
Long--Low (LL1: 19.5--38.0~\micron; LL2: 14.0--21.3~\micron) spectroscopic modes \citep{hou04}.  The widths of the SL and LL entrance
slits are 3\farcs7 and 10\farcs6, respectively.  For each mode, we obtained 6~s exposures at two positions along the slit.   We 
extracted and summed the spectra from the two slit positions to enhance the S/N and then subtracted the same spectra to establish the 
observational uncertainty above the normal shot noise and 5\% uncertainty in the absolute photometric calibration.  We then spliced 
the four spectral orders after removing their low-S/N edges, thereby obtaining a total wavelength range of 5.3--34~\micron.  Our measured
flux density of 42~mJy at 24~\micron\ matches very well the MIPS 24~\micron\ value reported by \citet{pla09}.  As before, we assumed a 
point-source aperture correction with the understanding that a secondary correction should be applied if the source is found to be 
extended (\S\ref{irmodel}).

\section{Analysis of Optical Scattered Light\label{optanal}}

\subsection{Surface Brightness\label{sbp}}

\begin{figure*}[t]
  \epsscale{1.8}\plotone{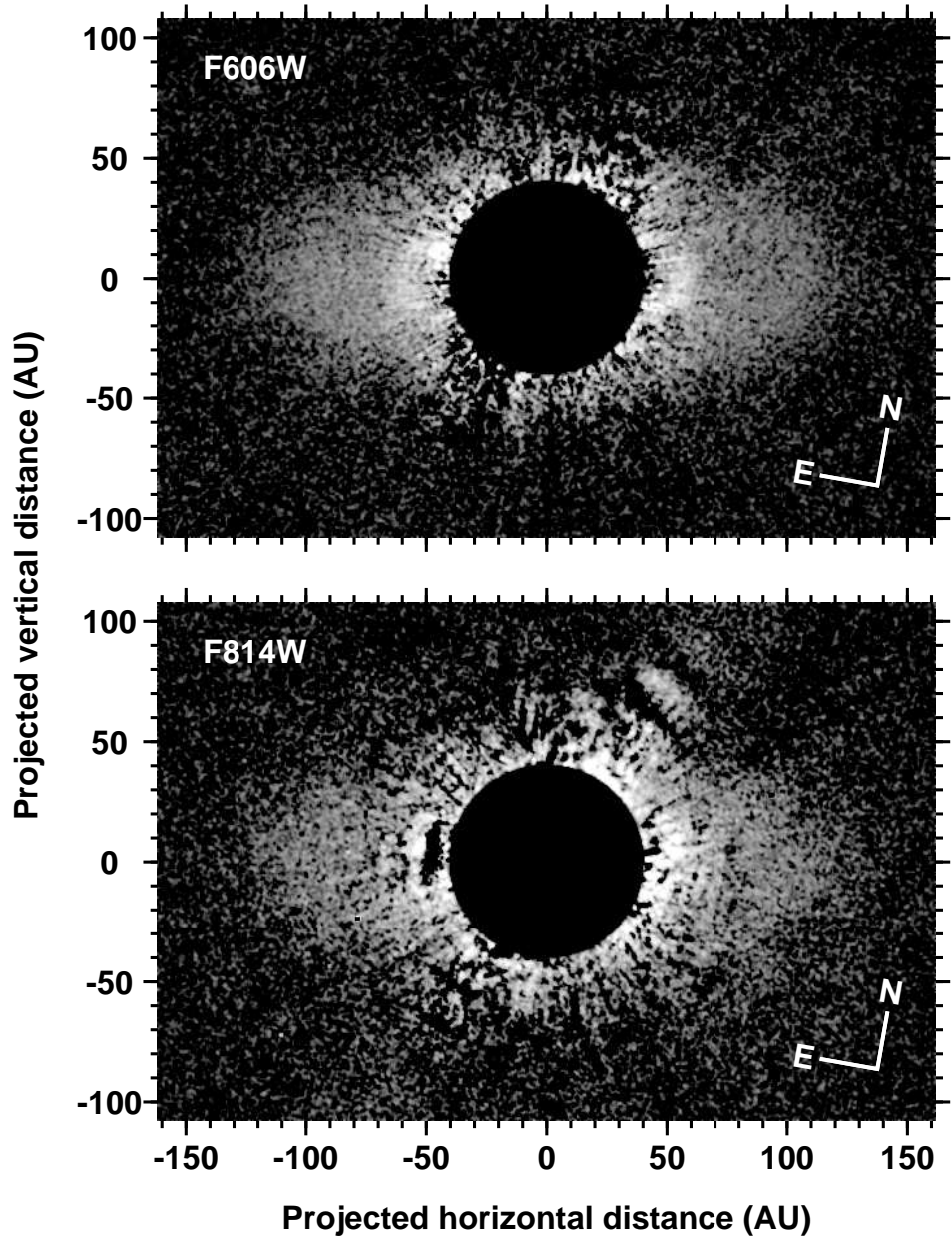}
   \caption{
	F606W and F814W images of HD~92945's disk after subtraction of the wide-angle halo.  Both images have been smoothed
        with a $3 \times 3$-pixel boxcar and rotated so that the projected major axis of the disk appears horizontal.  Artificial
	spots of radius 1\farcs9 mask the inner regions that are dominated by PSF-subtraction residuals.  The images are 
	displayed with logarithmic scaling.  
	}
  \label{subhalo}
\end{figure*}

The fully processed ACS/HRC images (Figure~\ref{subhalo}) suggest that HD~92945 is surrounded by an azimuthally symmetric disk of 
dust that is moderately inclined with respect to the line of sight and sharply confined to the region within $\sim 5$\farcs1 ($\sim 110$~AU)
of the star.   Any obvious azimuthal variations can be attributed either to geometric projection or to PSF-subtraction residuals.  
Such axisymmetry is rare within the present group of debris disks resolved in scattered light (Table~\ref{16disks}), which typically 
display asymmetries associated with perturbations from embedded planets or nearby stars (e.g., $\beta$~Pic and HD~141569A; 
\citealt{gol06,cla03}), interaction with the interstellar medium (e.g., HD~61005; \citealt{hin07}), and/or enhanced forward scattering 
by micron-sized dust grains (e.g., HD~15745; \citealt{kal07a}).  The apparent axisymmetry of HD~92945's disk permits us to examine the 
surface brightness profiles along the projected semimajor axes of the disk and to ascribe these profiles to the entire disk with 
reasonable confidence.

\begin{figure*}[t]
  \epsscale{2.0}\plotone{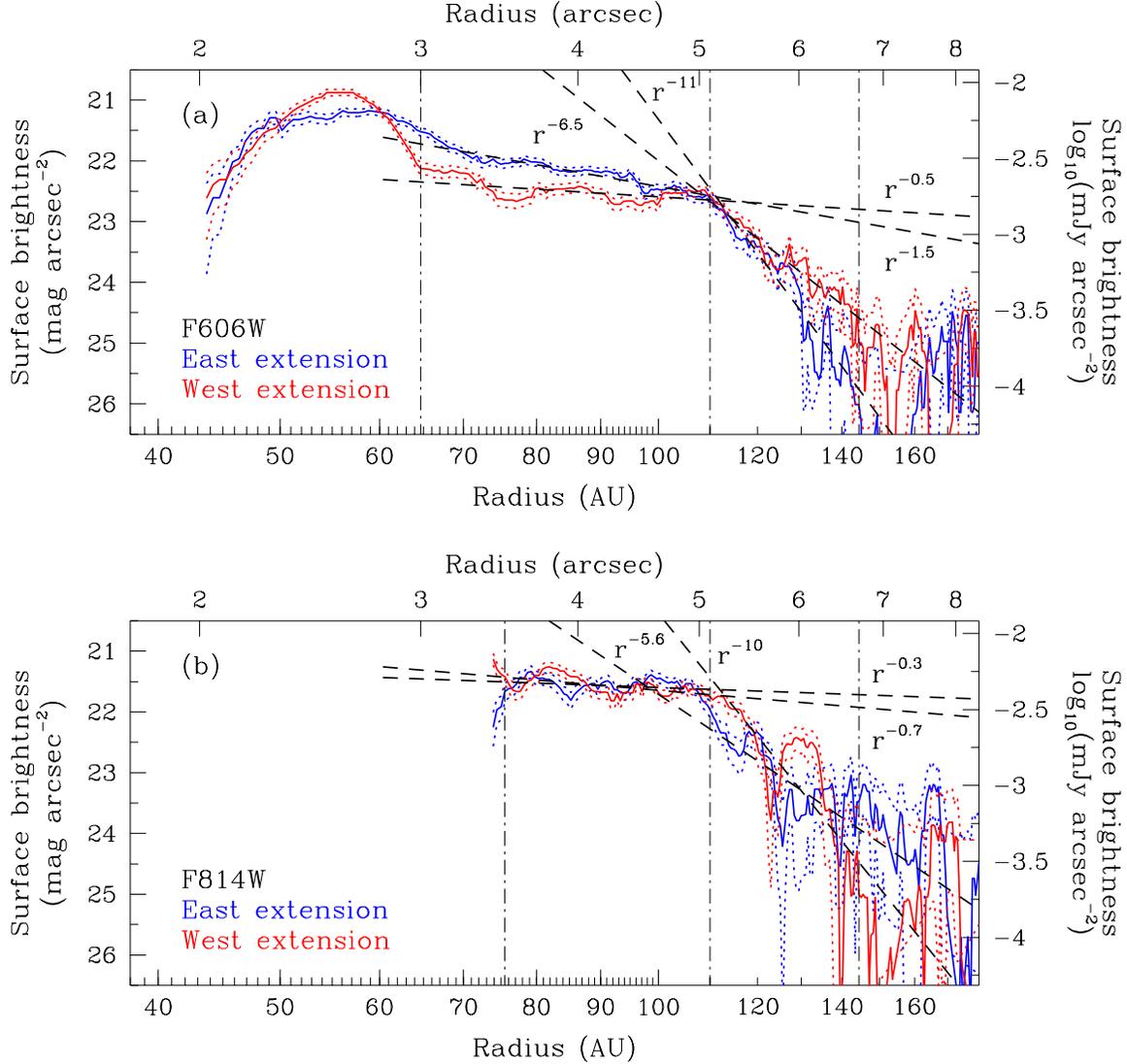}
   \caption{Surface-brightness profiles in the {\it (a)} F606W and {\it (b)} F814W bandpasses measured along each semimajor axis 
	of the disk as projected on the sky.  The solid curves show the radial profiles extracted from the fully processed HRC 
	images shown in Figure~\ref{subhalo} after smoothing the images with an $11 \times 11$ pixel boxcar.  The dotted curves show the same 
	profiles after addition and subtraction of the $1~\sigma$ error maps (see \S\ref{acspsfsub}) to the pre-smoothed images.  
	The heavy dashed lines represent radial power law fits to the regions 3\farcs0--5\farcs1 (F606W only), 3\farcs5--5\farcs1 
	(F814W only), and 5\farcs1--6\farcs7 (both F606W and F814W).  The vertical lines represent the boundaries of these regions.
	The surface brightnesses were calibrated by matching the observed brightness of HD~92945 in our unocculted HRC images to 
	synthetic photometry from a model K1~V spectrum ($T_{\rm eff} = 5000$~K, log $g = 4.5$; [M/H] = 0.0; \citealt{cas03}) 
	obtained with STScI's {\it synphot} package \citep{lai05}.  The synthetic magnitudes of HD~92945 itself are $m_{F606W} = 
	7.54$ and $m_{F814W} = 6.80$.
	}
  \label{sbew}
\end{figure*}

Figures~\ref{sbew}$a$ and \ref{sbew}$b$ show the F606W and F814W surface brightness profiles (and their $\pm 1~\sigma$ errors) along 
the opposing semimajor axes of the projected disk.   The profiles were extracted from the respective images shown in Figure~\ref{subhalo}
after smoothing the images with an $11 \times 11$ pixel boxcar.  The F606W profile has a credible inner radial limit of $\sim 2''$ ($\sim 43$~AU) 
because the two orientations used for the F606W images allow us to remove obvious PSF-subtraction artifacts near the occulting spot
(\S\ref{acspsfsub}).   However, the single orientation of the F814W image prevents credible measurement of the F814W profiles within
$\sim 3$\farcs4 of the star.  The F606W profiles are peaked between $2''$ and $3''$ (43--65~AU) from the star -- the region we call 
the inner ring -- with maxima of 21.0--21.5~mag~arcsec$^{-2}$.  In the region 3\farcs0--5\farcs1 (67--110~AU) from the star, the 
F606W and F814W profiles decline approximately as radial power laws, $r^{- \alpha}$, where $\alpha = 0.5$--1.5 for F606W and 
$\alpha = 0.3$--0.7 for F814W.  Beyond $r = 5$\farcs1 (110~AU), the profiles decrease so precipitously ($\alpha = 6$--11) that the 
disk is effectively truncated at that radius.

The $\pm 1~\sigma$ error profiles indicate no significant differences between the F814W profiles of the eastern and western sides 
of the projected disk, but the opposing F606W profiles differ by 2--$3~\sigma$ over most of the imaged region.   
These error profiles, which are obtained from our $1~\sigma$ error maps (\S\ref{acspsfsub}), reflect the combined uncertainties 
incurred from random errors (i.e., read and photon noise) and systematic errors from improperly scaled or aligned reference PSFs.
They do not include systematic errors stemming from any intrinsic but unquantifiable discrepancies between the coronagraphic PSFs of
HD~92945 and HD~100623 caused by differences in the stars' colors, {\it HST's} focus during the stars' observations, or the positions
of the stars behind the occulting spot.  The maximum difference of $\sim 0.5$~mag~arcsec$^{-2}$ between the opposing F606W profiles 
corresponds to a PSF-subtraction error of $\sim 20$\% at $r = 3''$, which is within the range of local residuals expected for focus
variations of a few microns typically associated with breathing or changes in {\it HST's} orientation with respect to the Sun 
\citep{kri00,kri03}.  Therefore, the F606W and F814W surface brightness profiles along the opposing semimajor axes are consistent 
with our initial assumption of an axisymmetric disk.

\subsection{Intrinsic Color\label{color}}

\begin{figure*}[t]
  \epsscale{2.0}\plotone{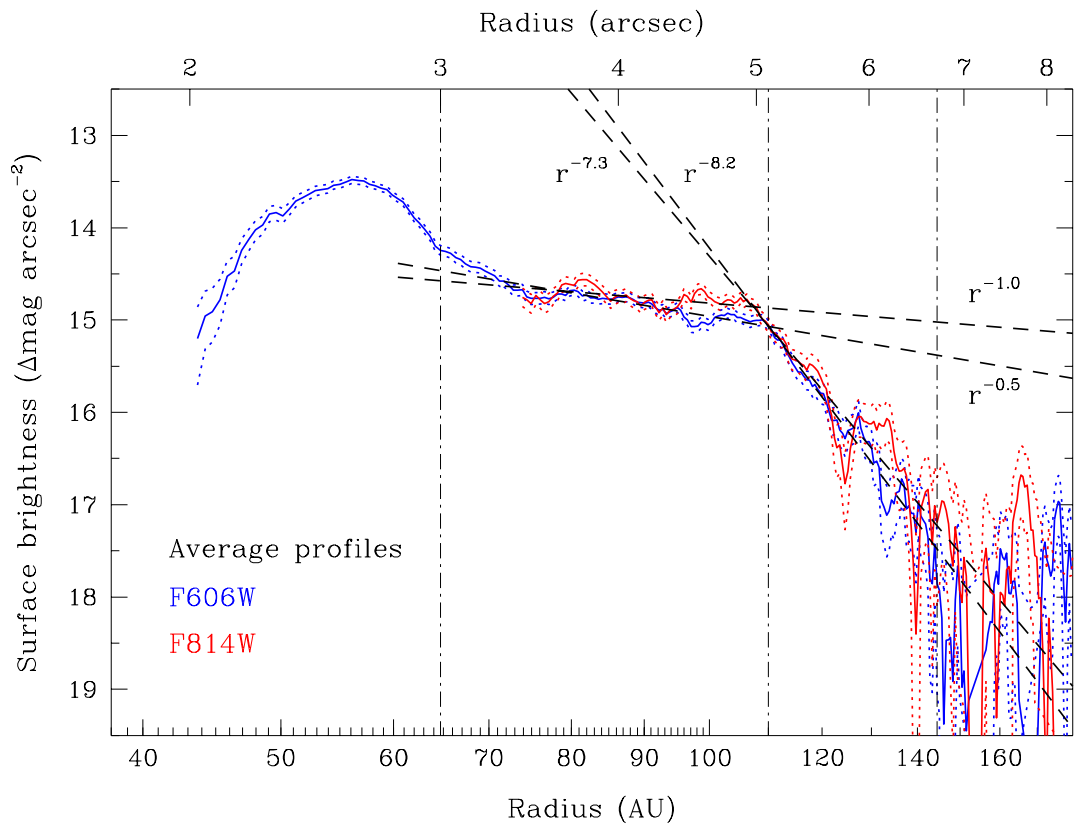}
   \caption{Average F606W {\it (blue curves)} and F814W {\it (red curves)} surface-brightness profiles measured along the semimajor 
	axes of the disk as projected on the sky.  All curves and lines conform to the descriptions given in Figure~\ref{sbew}.  The
	profiles have been normalized to the star's brightnesses in the respective bands, i.e., the ordinates are expressed in units
	of differential magnitude: $\Delta$mag arcsec$^{-2}$ = (mag arcsec$^{-2}$)$_{disk}$ -- mag$_{star}$.   The synthetic magnitudes
	of HD~92945 are $m_{F606W} = 7.54$ and $m_{F814W} = 6.80$.
	}
  \label{sbavg}
\end{figure*}

Figure~\ref{sbavg} shows the averages of the F606W and F814W surface brightness profiles extracted from both sides of the projected 
disk.  The profiles have been normalized to the measured brightnesses of the star in the respective bands, so that the ordinate now 
expresses the surface brightness of the disk as a differential magnitude: (mag arcsec$^{-2}$)$_{disk}$ -- mag$_{star}$.   The F606W
and F814W profiles are nearly coincident 3\farcs5--6\farcs7 from the star and can be represented by radial power laws with $\alpha 
\approx 0.75$ for $r = 3\farcs0$--5\farcs1 (65--110~AU) and $\alpha \approx 8$ for $r > 5\farcs1$.  These overlapping profiles indicate
that the disk has an intrinsic color of $m_{F606W} - m_{F814W} \approx 0$, i.e., the dust grains are neutral scatterers at red 
wavelengths.  This neutrality is unusual among the debris disks listed in Table~\ref{16disks}; only HD~32297's disk was previously 
reported to have intrinsic visible and near-infrared colors that are neither strongly blue nor red \citep{maw09}. 

The disk's neutral color at visible wavelengths suggests that the minimum size of the dust grains is at least a few microns for the 
expected varieties of grain compositions and porosities \citep{gol06}.  If so, then the near-infrared colors of the disk 
should not deviate significantly from gray.  Schneider et al.~(2011, in preparation) report a possible detection of HD~92945's disk 
using the coronagraphic imaging mode of {\it HST's} Near Infrared Camera and Multi-Object Spectrometer (NICMOS) in the F110W bandpass 
($\lambda_c = 1.1246$~\micron, $\Delta\lambda = 0.5261$~\micron; \citealt{via09}).\footnote{The NICMOS images of HD~92945 were obtained
in April 2005 as part of a coronagraphic survey of 57 circumstellar-disk candidates (GO Program 10177) that was planned independently of 
the ACS IDT program.}  The PSF-subtracted NICMOS image shows excess signal beyond $3''$ from the star in all directions, but prominent 
residuals from imperfect PSF subtraction obscure any morphology resembling the disk seen in our ACS images.  An azimuthal median 
profile of the unobscured circumstellar region indicates that the residual F110W surface brightnesses $3''$--$4''$ from the star are 
nearly an order of magnitude higher than those expected from our ACS images, assuming neutral scattering at visible and near-infrared
wavelengths.   If valid, the NICMOS results indicate that the disk exhibits a nonintuitive combination of gray colors at visible 
wavelengths and increased brightness around $\lambda \approx 1$~\micron.

\subsection{Surface Density of the Dust\label{surfdens}}

\begin{figure*}[t]
  \epsscale{2.0}\plotone{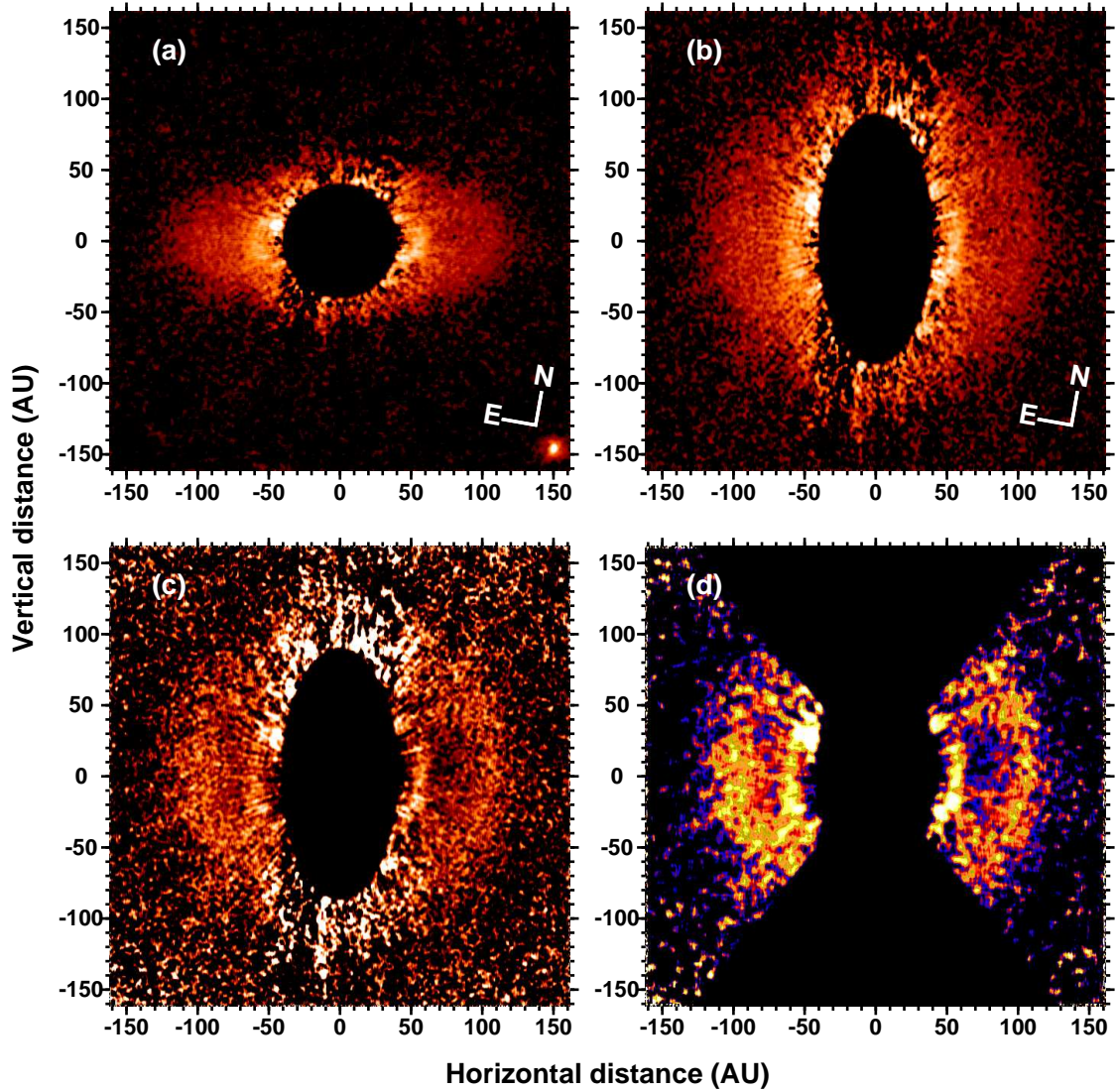}
   \caption{Conversion from surface-brightness image to scaled surface-density map ($\sigma\Sigma/4\pi$), assuming the dust has 
	homogeneous composition and scatters isotropically.  {\it (a):} Reproduction of F606W image shown in Figure~\ref{subhalo} using 
	linear color table and smoothed with a $5 \times 5$-pixel boxcar.  {\it (b):} F606W image after rotation of $\sim 62$\arcdeg\ 
	about its projected major axis so that the disk appears circular and coplanar with the sky.  {\it (c):} Deprojected image after 
	multiplying by $r^2$ to compensate for the geometric dilution of incident starlight.  {\it (d):} Same image shown in panel (c), 
	but smoothed with a $9 \times 9$-pixel boxcar and displayed with gamma-corrected color table.  The masked region has been 
	expanded to obscure areas containing large PSF-subtraction residuals.}
  \label{deproj}
\end{figure*}

The surface brightness $S$ of an optically and geometrically thin disk relative to the incident flux $F$ at a given angular radius 
$r$ from the star is 

\begin{equation}
\frac{4 \pi r^2 S}{F} = \sigma \Sigma,
\end{equation}

\noindent
where $\sigma$ and $\Sigma$ are the scattering cross section and surface density of the dust, respectively.  Assuming that the grains 
have homogeneous composition and scatter isotropically, we can map the surface density of dust in HD 92945's disk by scaling a 
deprojected version of the scattered-light image by $r^2$.  Figure~\ref{deproj} shows the sequential stages of the conversion of our 
F606W image to such a map.  Figures~\ref{deproj}{\it a} and \ref{deproj}{\it b} are false-color reproductions of our F606W image shown
in Figure~\ref{subhalo} before and after rotating the disk by $\sim 62$\arcdeg\ about its projected major axis so that the disk appears
circular and coplanar with the sky.  Figures~\ref{deproj}{\it c} and \ref{deproj}{\it d} show with different color tables and smoothing 
factors the deprojected image after multiplying by $r^2$ to compensate for the geometric dilution of incident starlight with angular
distance from the star.  These images represent the surface brightness of the dust scaled by an assumed constant value $\sigma/4\pi$.

Our scaled surface-density map reveals concentrations of grains not only in the previously described inner ring, but also along the 
outer edge of the disk defined by the sharp decline in the radial surface brightness profile at $r = 5\farcs1$ or 110~AU (\S\ref{sbp}).
These overdensities are apparent on both the eastern and western sides of the disk; they appear to be concentric rings of dust centered 
on the star.  The outer ring appears less azimuthally uniform on the eastern side, which may indicate nonisotropic scattering but more 
likely reflects lower S/N in the southeastern region of the image.   If the major axis of the projected disk is properly identified and 
the dust distribution is axisymmetric, then any evidence of forward scattering should be symmetric about the projected minor axis of the 
disk and not concentrated in its southeastern quadrant.

\subsection{Scattered Light Model\label{jekmodel}}

To obtain more quantitative constraints on the density distribution and scattering properties of the circumstellar dust, we applied 
the three-dimensional scattering model and nonlinear least-squares fitting code previously used to characterize the disk around 
AU~Mic \citep{kri05}.  Using the azimuthally averaged radial profile from Figure~\ref{deproj}{\it c} as a guide, the model 
simultaneously computes the scattering phase function, radial density profile, and inclination of the disk that best fits the observed
scattered-light image.  The model image also provides a better estimate of the ratio of integrated reflected light from the disk and 
emitted star light than can be obtained from coronagraphic images that have regions with large PSF-subtraction residuals.  Consequently, 
we modeled only our F606W image of the disk, which has better overall S/N and PSF subtraction than the F814W image.  

\begin{figure*}[t]
  \epsscale{2.0}\plotone{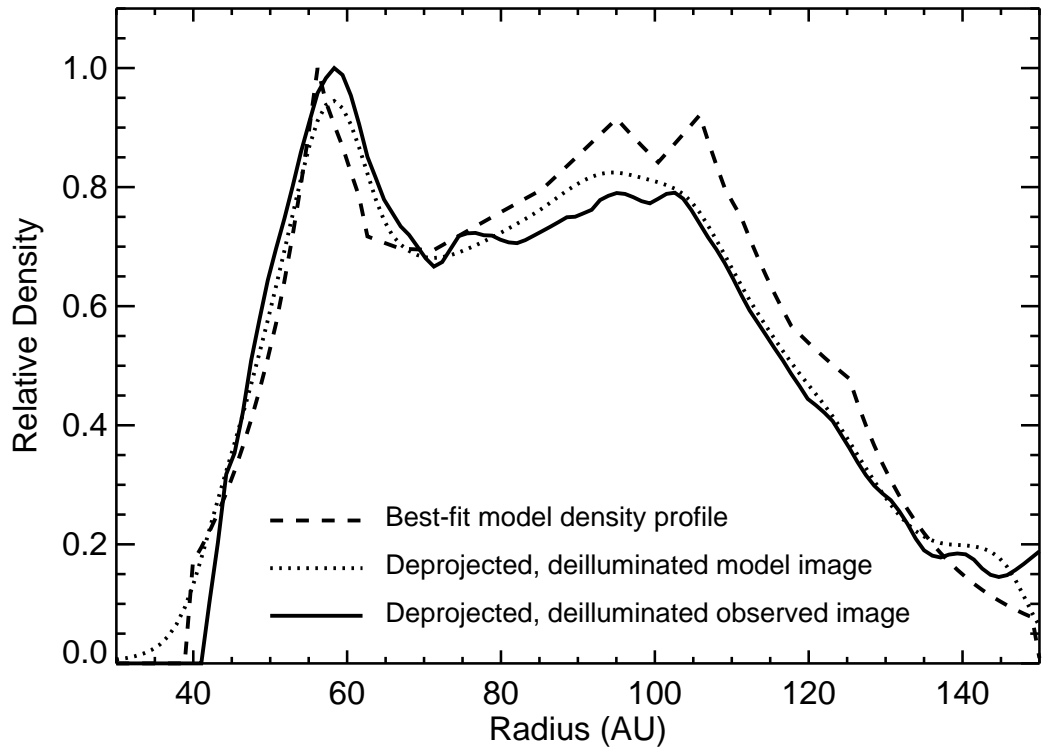}
   \caption{Radial profiles of scaled surface density obtained from observed and model F606W images.  The solid curve represents the 
	azimuthally averaged profile from the unmasked region of Figure~\ref{deproj}(d).  The dashed curve shows the 18-segment density 
	profile generated by our scattered-light model that best fits the F606W image of the disk, assuming a scale height of 0.5~AU.  
	The dotted curve is the equivalent of the solid curve for the best-fit model image (Figure~\ref{bestmodel}) after convolution with 
	a synthetic off-spot coronagraphic PSF obtained with the {\it HST} version of Tiny Tim \citep{kri04}.}
  \label{sdprof}
\end{figure*}

The F606W radial density profile used to guide the model is represented by the solid curve in Figure~\ref{sdprof}.  Because the model 
code is designed to produce best-fitting radial density profiles as a series of contiguous power-law functions, we allowed the code to 
fit up to 18 distinct power-law segments to approximate the unusually complex F606W profile of HD~92945.  These 18 segments were 
established for computational convenience; they are not associated with any physical attributes of the disk itself.  We set the inner 
radius of the model profile at 41~AU, as there appears to be a true clearing in the F606W image despite the large PSF-subtraction 
residuals near the star.  (The disk's surface brightness inside 41~AU certainly does not increase in a manner consistent with the 
steep increase associated with the outer edge of the inner ring.)  We also placed an outer limit of 150~AU on the model profile, in 
conformity with the largest distance at which the disk is confidently seen in our F606W image.

Because the plane of HD~92945's disk is inclined $\sim 28$\arcdeg\ from the line of sight (i.e., $\sim 28$\arcdeg\ from an ``edge-on''
presentation), our scattering model cannot constrain the vertical distribution of the dust as a function of radial distance from the 
star.  We can only derive a radial density distribution integrated along lines of sight across the disk.  Nevertheless, if we assume 
that the disk is vertically thin relative to its radial extent (as in the case of AU~Mic, whose disk is viewed edge-on), the 
models show that the vertical density profile does not strongly affect the appearance of a moderately inclined disk in scattered 
light.  We therefore adopted {\it a priori} for HD~92945 a Lorentzian vertical density profile similar to those observed in the edge-on 
disks around AU~Mic \citep{kri05} and $\beta$~Pic \citep{gol06}.   We also assumed a flat disk with a fixed scale height $z$ of either 
0.5~AU or 3.0~AU, which bracket the scale heights observed for AU~Mic's disk within $\sim 60$~AU of the star \citep{kri05}.

\begin{figure*}[t]
  \epsscale{1.5}\plotone{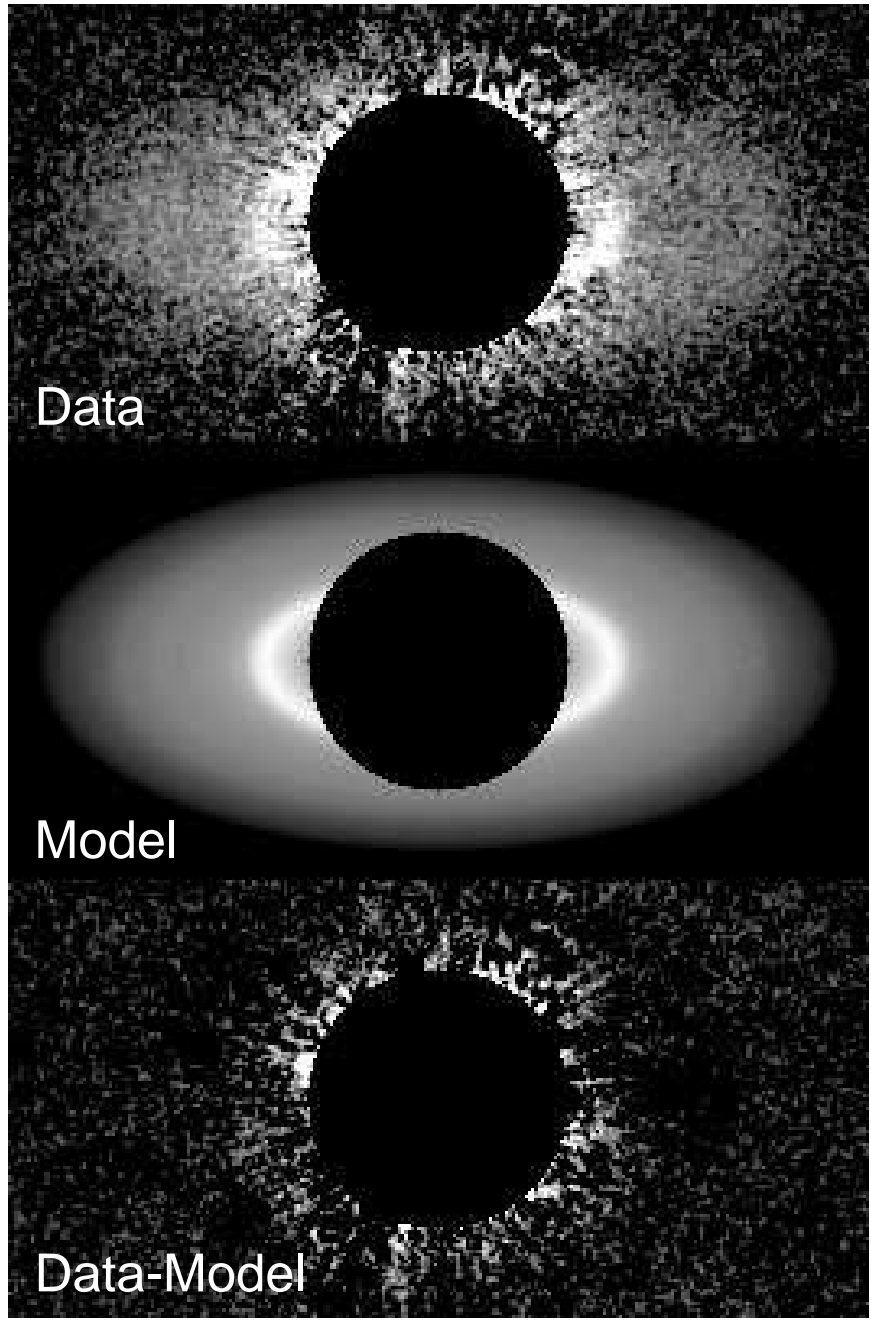}
  \vspace*{0.2in}
   \caption{Comparison of best-fit scattered-light model with observed F606W image of disk.  {\it (top:)} Reproduction of the final ACS 
	F606W image shown in Figure~\ref{subhalo}.  {\it (middle:)}  Model image generated for a constant scale height of 0.5~AU, after 
	convolution with a synthetic off-spot coronagraphic PSF obtained with the {\it HST} version of Tiny Tim \citep{kri04}.  {\it 
	(bottom:)}  Difference between observed and model images of the disk.}
  \label{bestmodel}
\end{figure*}

Having fixed the disk's inner and outer radii, thickness, and vertical density profile, we allowed our code to determine the radial 
density profile, intensity normalization, scattering phase function (as described by the asymmetry parameter $g$; \citealt{hen41}), 
and disk inclination that best fits the observed F606W image of HD~92945.  The model images were convolved with an ``off-spot'' 
coronagraphic HRC PSF appropriate for unocculted field sources \citep{kri00} before comparison with the actual F606W 
image.  Figure~\ref{bestmodel} shows the best-fit model for $z = 0.5$~AU; the corresponding model for $z = 3.0$~AU is effectively 
identical.  The dashed and dotted curves in Figure~\ref{sdprof} represent the radial-density profile of the best-fit model and the 
azimuthally averaged profile of the model's surface density map (i.e., the equivalent of Figure~\ref{deproj}c), respectively.  The 
models yield $g = 0.015 \pm 0.015$ for both vertical scale heights, which confirms our qualitative presumption of isotropic scattering.
The derived inclinations are 27\fdg4 and 27\fdg8 (relative to the line of sight) for $z = 0.5$~AU and 3.0~AU, respectively.  Finally,
both models yield an integrated F606W flux ratio of $(F_{disk}/F_*)_{F606W} = 6.9 \times 10^{-5}$.  

\section{Analysis of Thermal Emission\label{iranal}}

\subsection{Far-Infrared Spectral Energy Distribution\label{irsed}}

\begin{figure*}[t]
  \includegraphics[angle=270,scale=0.7]{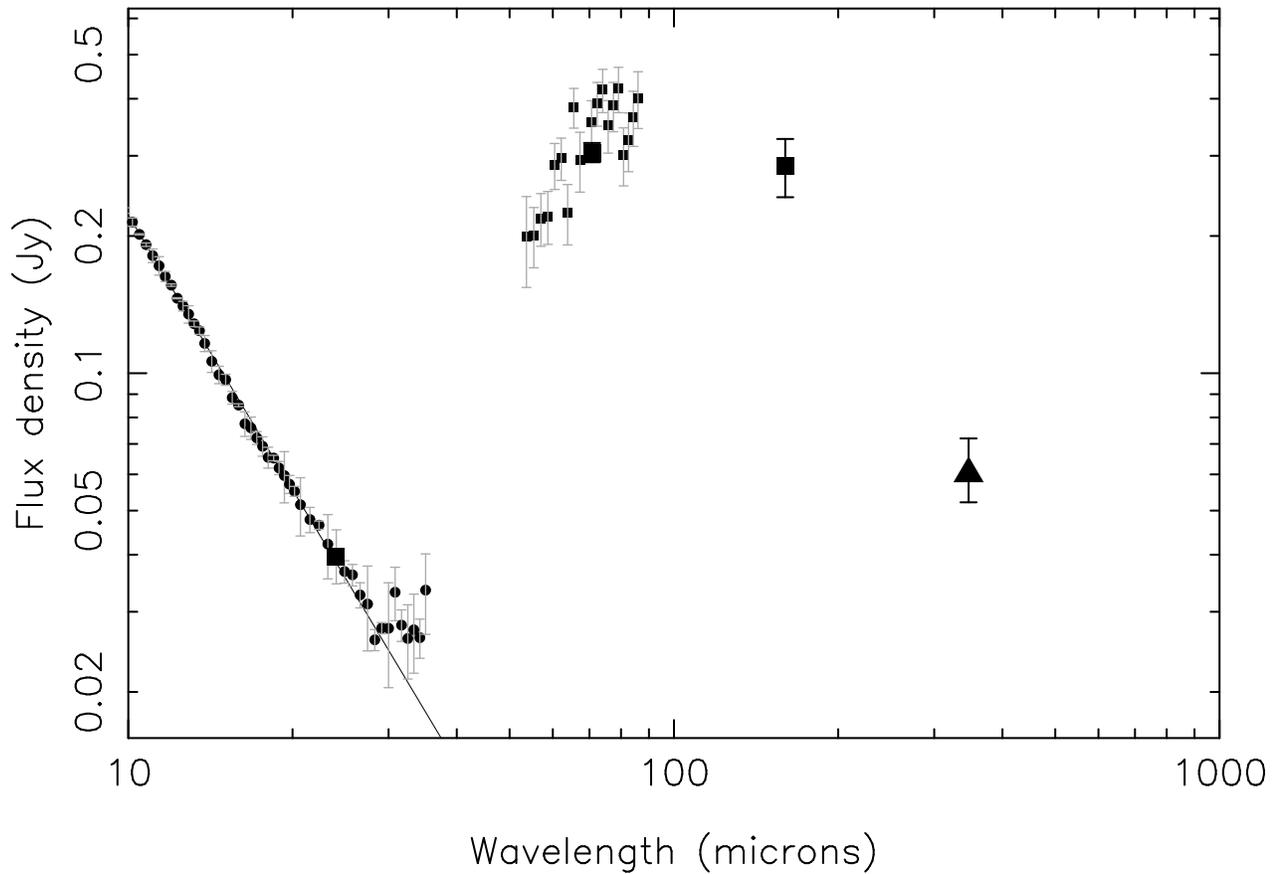}
   \caption{Spectral energy distribution of HD~92945's star and disk.  The small circles and squares show the IRS and MIPS SED data,
	respectively, after applying point-source aperture corrections for the finite slit widths.  The large squares represent the 24,
	70, and 160~\micron\ MIPS photometry reported in this paper and by \citet{che05}.  The triangle represents the CSO 
	350~\micron\ measurement previously reported by \citet{che05}.  The solid line shows the Rayleigh--Jeans extrapolation 
	from a $T_{\rm eff}= 5000$~K model atmosphere \citep{cas03} scaled to match the 2MASS photometry of HD~92945.}
  \label{sed}
\end{figure*}

Figure~\ref{sed} shows our MIPS and IRS data together with previously reported MIPS and ground-based submillimeter measurements of 
HD~92945.  The solid line represents the theoretical Rayleigh--Jeans profile extrapolated from a $T_{\rm eff}= 5000$~K model atmosphere 
\citep{cas03} normalized to the measured Two Micron All Sky Survey (2MASS; \citealt{skr97}) magnitude of $K_s = 5.66$.  The stellar 
photospheric emission detected with IRS closely tracks the theoretical Rayleigh--Jeans profile and dominates the total emission 
shortward of 30~\micron.  An infrared excess begins in the 30--35~\micron\ region of the IRS spectrum, rises steeply at the onset 
of MIPS SED coverage at 55~\micron, and then flattens across the 70--160~\micron\ region spanned by the MIPS broadband photometry.   
Undulations in the MIPS SED between 55 and 90~\micron\ are artifacts of the SED extraction process.  

Fits of elliptical gaussian profiles to the MIPS fine-scale 70~\micron\ and coarse-scale 160~\micron\ images reveal no significant 
source extension.  Our results constrain the full width of the 70~\micron\ source at half the maximum emission to $\lesssim 180$~AU.
These findings are consistent with the course-scale 70~\micron\ results of \citet{che05}.

\subsection{Modeling the Infrared SED\label{irmodel}}

We seek a model that reproduces our photometric and spectroscopic {\it Spitzer} observations of HD~92945, including an upper limit 
on the angular extent of the 70~\micron\ MIPS image.  To attain this model, we followed the strategy applied by \citet{kri10} to 
their {\it HST} and {\it Spitzer} observations of the dust ring surrounding HD~207129.  We initially assumed the following:

\begin{enumerate}
\item The radial distribution of dust obtained from our scattered-light model also delimits the dust responsible for the mid-infrared 
excess detected in the MIPS and IRS data.  
\item All dust is in local thermal equilibrium (LTE) with a star of luminosity $0.38~L_{\sun}$ and mass $0.77~M_{\sun}$.\footnote{The 
adopted values of luminosity and mass for HD~92945 are obtained from the NASA/IPAC/NExScI Star and Exoplanet Database 
(http://nsted.ipac.caltech.edu).}
\item The dust comprises astronomical silicate grains whose wavelength-dependent emissivities are calculated from Mie theory and the 
optical constants of \citet{lao93}.  
\item The grains have a size distribution, $dn/da \propto a^{-3.5}$, appropriate for a steady-state collisional cascade \citep{doh69}, 
but bounded between minimum and maximum radii $a_{min}$ and $a_{max}$.  
\end{enumerate}

We calculated model thermal images with $1''$ spatial resolution for each of 31 wavelengths spanning 10--850~\micron\ for this dust 
distribution and various combinations of vertical optical depth, $a_{min}$, and $a_{max}$.  Our initial value of $a_{min} = 
0.23$~\micron\ was based on the maximum radius ($a_{blow}$) of grains with mass density 2.5~g~cm$^{-3}$ that are immediately blown out 
from the disk by radiation pressure \citep{str06,pla09}.  We compared the total flux from each set of 31 models with the observed SED 
of HD~92945's excess infrared emission.  We combined appropriate subsets of the model images to simulate the MIPS 24, 70, and 
160~\micron\ broadband images.  To produce synthetic IRS and MIPS SED data, we convolved the model images with instrumental PSFs 
suitably windowed by the entrance slits and position angles used in our {\it Spitzer} observations.  Because the disk appears unresolved
in our broadband MIPS images, we assume that the losses of flux through the slits are identical to those of a point source.

The strongest constraint on $a_{min}$ is provided by the weak infrared excess seen in the IRS spectrum longward of 30~\micron.  Our 
initial model yielded $a_{min} = 3.5$~\micron, $a_{max} = 900$~\micron, and a vertical optical depth of 0.0027 for grains of size 
$a_{min}$ located at the inner edge of the disk ($r = 41$~AU).   Thus, $a_{min} \gg a_{blow}$.  Although these results are consistent 
with the observed upper limit of the 70~\micron\ source size, we had to steepen the grain-size distribution to $dn/da \propto a^{-3.7}$ 
to fit the 70~\micron\ and 160~\micron\ photometry simultaneously.  Even with this adjustment, the initial model 
overestimated the 350~\micron\ flux density by a factor of 1.7.  

To address this problem, we postulated that the large grains are not evenly distributed throughout the observed disk but are confined 
to the inner ring at 41--63~AU.  We also surmised that the outer disk (63--150~AU) consists of smaller particles that have been ejected 
from the inner ring by some combination of collisions, stellar wind, radiation pressure, and/or gravitational scattering by planetary 
bodies.  Similar structures have been inferred for the disks surrounding AU~Mic and $\beta$~Pic \citep{str06}, HD~61005 \citep{man09},
Fomalhaut \citep{chi09}, and Vega \citep{mul10}.  We thus adopted a two-component dust model for HD~92945's disk.  We maintained the 
grain-size distribution and surface-density profile derived from the scattered-light images, but assumed that $a_{max}= 900$~\micron\ 
within the ring and $a_{max}= 50$~\micron\ elsewhere in the disk.  This change reduced the total number of large grains in the disk and
brought the model into agreement with the 350~\micron\ measurement.   

While this two-component model provides a good fit to the MIPS and IRS data, it is inconsistent with the observed surface brightness 
of the disk in scattered light.  The average albedo of the dust is 

\begin{equation}
\langle\omega\rangle \approx \frac{F_{scat}}{F_{scat} + F_{emit}},
\end{equation}

\noindent
where $F_{scat}$ and $F_{emit}$ are the fractions of bolometric stellar flux that are scattered and emitted by the dust, respectively.
According to Mie theory, astronomical silicate grains with $a > 1$~\micron\ should have $\langle\omega\rangle \approx 0.55$ from visible 
to near-infrared wavelengths \citep{vos05}.  We used this value in the LTE calculations intrinsic to our SED model.  However, if we assume 
from our ACS F606W and F814W images that the disk scatters neutrally at all wavelengths that significantly heat the dust (notwithstanding
the NICMOS results of Schneider et al.\ 2011, in preparation), then our scattered-light model implies $F_{scat} = 6.9 \times 10^{-5}$ 
(\S\ref{jekmodel}).  We also derive $F_{emit} = 6.0 \times 10^{-4}$ from the measured infrared SED (Figure~\ref{sed}), which supersedes 
previous estimates of $L_{IR}/L_*$ reported by \citet{che05} and \citet{pla09}.  We therefore obtain $\langle\omega\rangle \approx 0.10$, 
which is $\sim 5$ times smaller than the predicted Mie value.  As noted before for HD~207129's disk \citep{kri10}, simple Mie grains are 
inconsistent with the combined visible and far-infrared observations of the disk surrounding HD~92945. 

\begin{figure*}
\includegraphics[angle=270,scale=0.7]{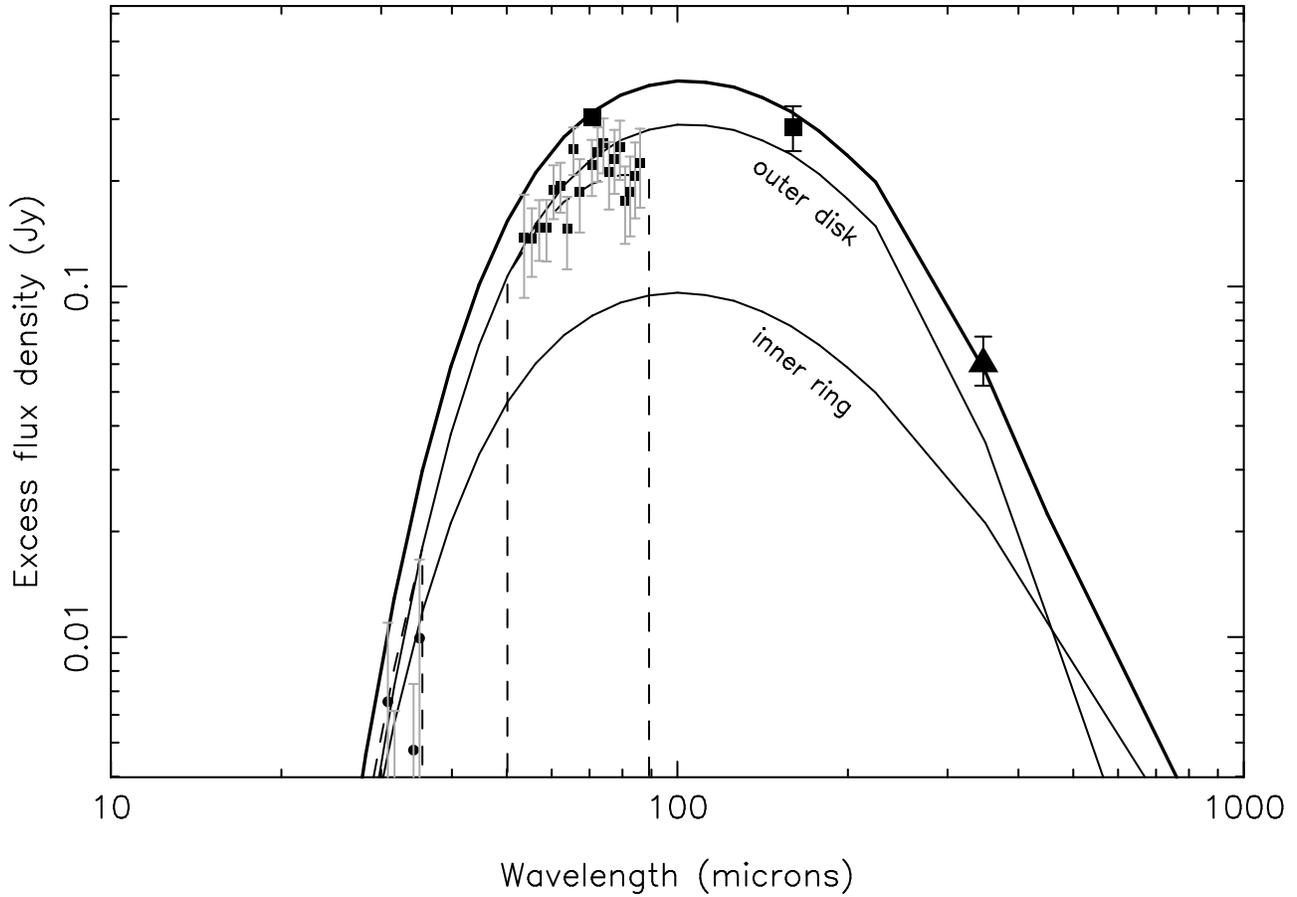}
\caption{Comparison of thermal models of HD~92945's disk (\S4.2) with observed excess infrared flux after removal of the Rayleigh--Jeans fit to 
	the stellar photosphere.  An average albedo of $\langle\omega\rangle = 0.10$ is assumed.  The measured fluxes are represented 
	with the same symbols described in Figure~\ref{sed}, but the IRS and MIPS SED data are no longer corrected for point-source flux
	losses due to the finite slit widths.   The thick solid line shows our preferred two-component disk model; the thin solid lines 
	reflect the contributions from the inner ring and the outer disk.   The dashed lines show the disk model windowed by the IRS and 
	MIPS SED slits.}
	\label{sedmodel}
\end{figure*}

We produced another two-component model using $\langle\omega\rangle \approx 0.10$ but retaining the original Mie value for the grain 
absorption efficiency, $Q_{abs}$.  In other words, we retained the emissive properties of astronomical silicate grains but modified their 
reflectivity so they are essentially black.  To compensate for the effect of this adjustment on the short-wavelength end of the SED, 
we increased $a_{min}$ to 4.5~\micron\ because larger, less reflective grains achieve the same equilibrium temperature as smaller, more 
reflective grains at a given distance from the star.  Figure~\ref{sedmodel} shows our final model, which best matches the observed SED
and surface brightness of the disk if the sources of the scattered and emitted light are cospatial.  The component SEDs show that the 
outer disk is the dominant contributor to the observed 70, 160, and 350~\micron\ emission and that the larger grains in the inner ring 
should become the dominant emitters longward of $\sim 400$~\micron.  

The constraints placed on $dn/da$, $a_{min}$, and $a_{max}$ by our final model are unique for the assumed emissive properties of compact 
astrosilicate grains, but other constraints may be obtained for different assumptions about these properties.  Supplemental photometry
at 450 and 850~\micron\ would be a useful check of our values of $a_{max}$ in both the inner ring and outer disk.  We can however 
exclude blackbody grains from consideration, as they would be too cold at the radii delimited by the scattered-light images to produce the 
observed SED.  Moreover, the disk is too extensive to be characterized by a single blackbody temperature.  Any resemblance of our SED model
longward of 160~\micron\ (Figure~\ref{sedmodel}) to the Rayleigh--Jeans tail of a single Planck distribution is coincidental.  

\subsection{Integrated Dust Mass\label{dustmass}}

Combining our final values of $dn/da$, $a_{min}$, and $a_{max}$ for each dust component with the surface-density profile from our 
scattered-light model, we obtain an integrated dust mass of $5.5 \times 10^{24}$~g ($\sim 0.001~M_{\earth}$) for compact grains of pure 
astronomical silicate.  Although $a_{max}$ is 18 times larger in the inner ring than the outer disk, the inner ring is 4 times less 
massive than the outer disk.  This disparity is caused by the relatively large surface area of the outer disk and the steeply declining 
grain-size distribution throughout both the inner ring and outer disk.  Our estimated mass of $1.1 \times 10^{24}$~g for grains smaller
than 900~\micron\ in the inner ring is $\sim 1$\% of the computed mass of Fomalhaut's ring \citep{hol98,kal05a},\footnote{\citet{kal05a}
computed masses of (0.7--1.5) $\times 10^{26}$~g for Fomalhaut's ring assuming albedos of 0.1--0.05, which conform to our measured value
of $\langle\omega\rangle = 0.10$ for HD~92945.  For the extreme case of $\langle\omega\rangle = 1$, \citet{kal05a} computed a mass of 
$7.4 \times 10^{24}$~g, which is comparable with our estimate for HD~92945's inner ring.} and $\sim 0.05$\% of the predicted mass of 
the ring around HR~4796A \citep{kla00}.

Our integrated dust mass is $\sim 60$ times larger than the value obtained by \citet{pla09} from early MIPS 24 and 70~\micron\ 
photometry of HD~92945.  This discrepancy is due to prior assumptions that the dust lies within a thin shell of radius 23~AU, $dn/da \propto 
a^{-3.5}$, and $\langle a \rangle \approx 0.25$~\micron.  None of these assumptions is upheld by our scattered-light images and thermal model.  
For comparative purposes, we applied the same formula used by \citet{pla09} to compute the dust mass and obtained a value that is only 20\% 
larger than the one derived from our observed surface-density profile.   Therefore, our revised dust mass of $\sim 0.001~M_{\earth}$ supersedes
those previously reported by \citet{che05} and \citet{pla09}.

\section{Discussion\label{discussion}}

\subsection{Dust Properties\label{dustprop}}

The dust's neutral color and negligible emission at 24~\micron\ indicate that the collisional cascade is suppressed below radii of 
several microns, which is an order of magnitude larger than $a_{blow}$.   This truncation of the grain-size distribution is 
counterintuitive for a star of subsolar mass and luminosity, especially as the dust's isotropic scattering and low albedo indicate a 
large population of submicron-sized grains \citep{vos05}.  This contradiction suggests that our initial assumptions about the location, 
composition and/or size distribution of the thermally emitting dust should be reviewed.\\

\subsubsection{Radial Distribution\label{dustrad}}

Our invocation of two dust components with common $a_{min}$ but different $a_{max}$ to explain the disk's neutral color and infrared 
SED is based on the assumption that the dust grains responsible for the scattered light and the infrared excess are cospatial.  
Although the inner ring seen in our coronagraphic F606W image suggests that the dust is created in a coincident planetesimal belt, the
presence of dust closer to the star cannot be ruled out by our coronagraphic images alone.  If such dust is present and in LTE with the 
stellar radiation field, it must have $a_{min} > 4.5$~\micron\ (our derived value for the inner ring) or a large albedo to avoid a 
detectable 24~\micron\ excess.  The former option worsens the discrepancy between the large $a_{min}$ needed to fit the infrared 
SED and the small $a_{blow}$ for a young K dwarf like HD~92945.  The latter option counterintuitively suggests that conditions favorable 
for icy grains increase within 40~AU of the star rather than beyond the expected snow line at $\sim 2$~AU \citep{ken08}.  Thus, neither 
option favors the presence of much dust between the star and the inner ring.

Conversely, if much dust exists beyond the apparent outer limit of the scattered-light disk ($r \approx 5\farcs1$ or 110~AU), then its 
albedo must be nearly zero to conform with the observed surface brightness profiles (Figure~\ref{sbavg}).   Moreover, $dn/da$ 
must be much steeper than the traditional $a^{-3.5}$ power law \citep{doh69} to account for both the observed SED longward 
of 160~\micron\ and the unresolved 70~\micron\ image of the disk.   Such a small albedo again counters the expected formation of icy 
mantles beyond the snow line, and it implies a corresponding increase in emissivity for grains larger than $\sim~0.1$~\micron.   Such small 
grains would be efficiently expelled from the disk by radiation pressure.   Collectively, these arguments support our assumption that 
all the thermally emitting dust is confined within the angular limits of the visible scattered light seen Figure~\ref{subhalo}.

\subsubsection{Composition and Porosity\label{dustcomp}}

Recent theoretical studies of other debris disks have frequently invoked highly porous aggregate grains to explain the observed 
scattered light and/or thermal emission from the dust \citep{li98,li03a,li03b,llb03,gra07,fit07,she09}.   Models of the optical 
properties of porous grains have yielded diverse and sometimes contradictory results, so we defer a detailed model of HD~92945's 
disk using porous aggregate grains to other investigations.  For now, we qualitatively assess the substitution of such grains for 
the compact grains in our scattered-light and thermal models.  

\citet{vos05} showed that the Henyey--Greenstein scattering asymmetry parameter, $g$, of moderately porous grains increases by factors 
of 1--1.25 over that of compact grains for $a = 0.5$--2.5~\micron\ and $V$-band wavelengths.  \citet{she09} showed that the degree of 
porosity has little effect on the scattering phase functions of grains at wavelengths $\lambda \approx 2 \pi a$.  Consequently, neither 
compact nor porous micron-sized grains account for the isotropic scattering observed from HD~92945's disk.  The effect of porosity on 
the albedo is unclear, however, as \citet{hag90} and \citet{vos05} oppositely predict that highly porous grains have lower and higher 
albedos, respectively, than compact grains of similar composition.  

The effect of porosity on the sizes of grains that succumb to radiation pressure is also ambiguous because $a_{blow}$ also depends 
on the composition of the grains.   \citet{muk92} showed that the ratio $\beta$ of the forces of radiation and gravity on a grain 
decreases with increasing porosity for $a \lesssim 1$~\micron, regardless of grain composition.  \citet{sai03} confirmed this result
and emphasized that the decreasing $\beta$ was limited to ultraviolet and visible wavelengths.  \citet{muk92} also showed that $\beta$
increases with porosity for absorptive grains (e.g., magnetite and graphite) with $a \gtrsim 1$~\micron, and it becomes nearly 
independent of grain size at the highest porosities.  On the other hand, dielectric grains (e.g. astronomical silicate) with $a 
\lesssim 10$~\micron\ experience decreasing $\beta$ as porosity increases.  \citet{koh07} determined that $\beta$ increases rapidly 
with increasing porosity for both graphite and silicate grains with $a \gtrsim 10$~\micron.  Collectively, these results indicate that 
high porosity of grains in HD~92945's disk does not explain the $a_{min} \approx 4.5$~\micron\ obtained from our silicate-based thermal
model.  However, high porosity may be responsible for such a large value of $a_{min}$ if the dust is mostly composed of more absorptive
materials like graphite.

\subsubsection{Size Distribution\label{sizedist}}

Our need to steepen HD~92945's grain-size distribution from the traditional $a^{-3.5}$ representation \citep{doh69} to reproduce the 
observed SED is consistent with thermal models of the disks around AU~Mic \citep{str06} and HD~207129 \citep{kri10}, which, like our 
model, employ compact spherical grains.  On the other hand, models that employ fluffy porous grains, like those developed for the 
disks around HR~4796A \citep{li03a}, HD~141569A \citep{li03b}, and $\epsilon$~Eridani \citep{llb03}, require grain-size distributions
that are significantly shallower than $a^{-3.5}$.  Recent models of collisional cascades in the presence of radiation pressure have 
shown that the rapid loss of grains with $a < a_{blow}$ creates a wave in the size distribution of the remaining grains \citep{the03,
krv06,the07}, so a single power-law representation may be inappropriate regardless of porosity or composition.  \citet{the07} noted 
that the wave becomes sufficiently damped for $a \gtrsim 100~a_{blow}$ that the size distribution can be approximated by $a^{-3.7}$, 
which matches our best-fitting thermal model of HD~92945.  This agreement is surprising because our model suggests that most grains 
in the outer disk, which is the dominant contributor to the observed SED at $\lambda \lesssim 400$~\micron\ (Figure~\ref{sedmodel}),
are smaller than $100~a_{blow} = 23$~\micron\ and should therefore lie in the wavy part of the distribution.  Our result may indicate 
that the wave is actually smaller and/or less extensive than predicted by \citet{the07} and that the size distribution follows the 
$a^{-3.7}$ power law for grains as small as $10~a_{blow}$ (A.~G\'{a}sp\'{a}r, personal communication).  

\citet{the03} noted that the wavy structure in the size distribution is more pronounced for materials that are prone to cratering
rather than catastrophic fragmentation.   If so, then softer and more absorptive grains like graphite may not easily account for both
the large $a_{min} \approx 4.5$~\micron\ (\S5.1.2) and the monotonic $a^{-3.7}$ size distribution obtained from our thermal model. 
Of course, our model is based on the common {\it a priori} assumption of a power-law distribution without any regard to goodness-of-fit,
so we cannot discount any possible compositions of the dust without an exhaustive investigation of alternative size distributions.\\

\subsubsection{Effects of Stellar Wind\label{wind}}

\citet{pla05,pla09} studied the tangential and radial contributions of corpuscular stellar wind to the removal of grains in the debris 
disks of late-type dwarfs.  They found that the tangential component (or ``corpuscular drag'') is more important than Poynting--Robertson
drag for K and M dwarfs whose mass-loss rates from stellar wind ($\dot{M}_{sw}$) are similar to the solar rate.   \citet{str06} obtained 
$\dot{M}_{sw} \lesssim 10~\dot{M}_{\sun}$ for AU~Mic from their scattered light and thermal model of its disk, but they concluded that 
corpuscular drag is not a significant mechanism of grain removal from the disk because the grains are more quickly destroyed by mutual 
collisions within the dust's ``birth ring.''  \citet{pla09} confirmed that the collisional lifetime of grains in the birth ring is 
$\sim 10^4$ times smaller than the timescale for removal by corpuscular drag, so stellar wind does not presently contribute to the 
evolution of AU~Mic's disk.

\citet{woo05} developed an empirical relationship between X-ray luminosity and $\dot{M}_{sw}$ for G, K, and M dwarfs with X-ray 
luminosities $\lesssim 8 \times 10^5$~erg~cm$^{-2}$~s$^{-1}$ and ages $\lesssim 700$~Myr.  Although HD~92945's X-ray luminosity
of $1.4 \times 10^6$~erg~cm$^{-2}$~s$^{-1}$ is formally beyond the applicable range of this relation, we nonetheless apply the relation
to determine whether stellar wind may be a contributing factor to the large $a_{min}$ observed in HD~92945's disk.  We find that 
$\dot{M}_{sw} \approx 100~\dot{M}_{\sun}$, which implies that the stellar wind augments the radiation pressure on the 
grains by only $\sim 8$\% \citep{pla09}.  In other words, HD~92945's radiation and corpuscular wind together yield $a_{blow} \approx 
0.25$~\micron, which is 18 times smaller than $a_{min}$ obtained from our thermal model.

For $a_{blow}$ to be consistent with $a_{min}$, HD~92945's mass-loss rate would have to be an implausible $2 \times 10^4~\dot{M}_{\sun}$. 
However, if the stellar wind velocity was larger than $\upsilon_{sw}/c = 10^{-3}$ assumed by \citet{pla09}, then the required $\dot{M}_{sw}$
could be significantly reduced.  This possibility is unlikely, however, because $\upsilon_{sw}$ is approximately equal to the escape 
velocity, which is approximately constant for stars on the lower main sequence.  We therefore conclude that neither corpuscular drag nor 
blow-out is responsible for the $a_{min}$ derived for HD~92945's inner ring. 

\subsection{Disk Morphology \label{diskmorph}}

The other known neutrally scattering debris disk surrounds the A star HD~32297 \citep{maw09} and presents a nearly edge-on and highly 
asymmetric appearance that has alternately been attributed to collisional interaction with the interstellar medium \citep{kal05b}, the 
destruction of a large planetesimal \citep{gri07}, or the resonant trapping of dust by an inner planet \citep{man08}.  Although 
HD~92945's disk shows no significant asymmetry along its projected major axis, its morphology may represent an azimuthally smoother 
remnant of collisions within one or more planetesimal belts.  In fact, the combination of an inner ring surrounded by a diffuse disk 
conforms very well to structures predicted by a variety of dynamical models that explore the evolution of debris disks as embedded 
planets form and/or migrate \citep{ken02,ken04,wya03, wya06,the08}, or as the dust alone migrates in the face of radiation pressure 
and gas drag \citep{kla00,tak01}.  HD~92945's age \citep{pla09} and undetected H$_2$ emission \citep{ing09} suggest that the disk is 
largely depleted of gas, so we discount gas--dust coupling as a viable cause of the disk's ringed structure.

\subsubsection{Comparison with Dynamical Models\label{modcomp}}

The surface brightness and density maps of \linebreak HD~92945's disk (Figure~\ref{deproj}) strikingly resemble those produced by \citet{ken04}
and \citet{wya06} from their dynamical models of dust created from collisions of planetesimals confined to rings or planetary 
resonances.  \citet{ken04} modeled the scenario in which the collisions occur in an expanding ring associated with an outwardly 
propagating wave of planet formation.  At any given epoch, this scenario conforms to the ``birth ring'' concept of \citet{str06}.  
\citet{wya06} considered an alternative scenario in which the colliding planetesimals are trapped in a gravitational resonance of 
an outwardly migrating planet.  The resulting dust grains either remain in or migrate from resonance according to their size (or, 
more accurately, their associated value of $\beta$).  In this model, Kuiper Belt grains with $0.008 \lesssim \beta \lesssim 0.5$ 
would fall out of resonance with a migrating Neptune but remain bound to the Sun in increasingly eccentric and axisymmetric orbits 
as they are scattered by multiple close encounters with the planet.

Whereas HD~92945's inner ring may be readily described by such dynamical models, the precipitous decline in its surface brightness 
beyond 110~AU is more problematic.   For example, \citet{ken04} initially considered a $3~M_{\sun}$ star with a quiescent disk having
a surface-density profile $\propto r^{-1.5}$ at $r = 30$--150~AU, which after several hundred Myr remains smoothly asymptotic at large
radii except when a ring, gap, or shadow passes through the region.  Changes in the the initial parameters of the system affected 
the time scales but not the outcomes of the disk evolution.  To reproduce HD~92945's double-peaked surface-density profile 
(Figure~\ref{sdprof}), this model requires two concurrent waves of planet formation at the inner and outer edges of the imaged disk.
\citet{ken04} showed that a close encounter with a passing star can also initiate a wave of planet formation at the inner edge of the 
disk, but we identify no such fly-by candidates in our {\it HST} and {\it Spitzer} images.\footnote{Five field sources are seen in the 
ACS/HRC F606W images obtained on UT~2004 December~1 and UT~2005 July~12, but all have motion relative to HD~92945 that is consistent 
with a distant background object.  No field sources within $2'$ of HD~92945 are seen in the broadband MIPS images.}  Therefore, the 
model of \citet{ken04} is probably insufficient to explain the apparently sharp boundaries of HD~92945's disk.

The resonant-dust model of \citet{wya06} accommodates sharp inner and outer edges of the disk if the grains have values of $\beta$ 
between those necessary for escape from the gravitational resonance and radiative expulsion from the disk.  The former limit depends
on the planet's mass, so we can assess whether the different values of $a_{max}$ obtained from our thermal model of the inner ring 
and outer disk (\S\ref{irmodel}) are consistent with the presence of a migrating planet between the star and the inner ring.   If 
we crudely extrapolate plots of $\beta(a)$ computed for a variety of solar-system grains with $a < 10~\micron$ \citep{bur79} and 
apply them to the dust surrounding HD~92945, then our value of $a_{max} = 50$~\micron\ in the outer disk indicates that $\beta
\gtrsim 0.004$ for compact grains that leave the resonance.  For a $0.77~M_{\sun}$ star like HD~92945, this lower limit of $\beta$ 
is consistent with a migrating planet of mass $\sim 3~M_{\earth}$ \citep{wya06}.  This scenario implies that larger grains in the 
disk -- like those confined by our thermal model to the inner ring -- librate resonantly over a broad but finite range of azimuths.

Returning to the birth-ring scenario, \citet{the08} found that the surface brightness of an optically thin dust disk of mass $\sim 
0.1~M_{\earth}$ achieves a collisionally steady-state profile $\propto r^{-3.5}$ with no sharp outer edge.  The smooth decline at 
large distances is mainly caused by the forced outward migration of small grains by radiation pressure.   They also found that if 
either the migration or the production of small grains were somehow inhibited, then the disk would retain the sharp edge initially defined 
by the planetesimal belt.  The former possibility does not apply to HD~92945 because its disk has very little mass (\S\ref{dustmass})
and is optically thin, but the latter one may be relevant to both the morphology and large $a_{min}$ of the disk if it is 
dynamically cold.  \citet{the08} determined that the low eccentricities of bodies in dynamically cold disks not only reduce the 
production of small grains from destructive collisions of larger bodies, but also increase the destruction of small grains because 
the grains are more likely to collide with larger bodies as they are radiatively expelled from the disk.   Although the potential 
link between a cold disk and large $a_{min}$ is intriguing, the low eccentricities needed for such a condition conflict with the high
eccentricities of grains in the outer disk (caused perhaps by repeated encounters with a migrating planet) that likely produce the
double-peaked surface-density profiles observed for HD~92945 (Figure~\ref{sdprof}) and modeled by \citet{wya06}.

\subsubsection{Effect of Disk Inclination\label{incline}}

The intermediate inclination of HD~92945's disk makes our scattered-light model (\S\ref{jekmodel}) insensitive to the disk's vertical
thickness and its possible radial dependence.  We are therefore unable to constrain the scale height of the disk, which is an indicator
of its dynamical temperature and a means for assessing the relevance of the dynamical models just described.  Moreover, the disk's 
inclination inhibits the detection of small-scale clumps or perturbations like those seen in the disks of AU~Mic \citep{kri05} and
$\beta$~Pic \citep{gol06}, which are viewed along their edges through extensive columns of dust.   Consequently, we are unable to 
ascribe the apparent axisymmetry of the dust to the dominance of radiation pressure in a quiescent disk, a temporary lull in the 
stochastic collisions of planetesimals, or some other process.  Resolved images of the disk at far-infrared or submillimeter 
wavelengths would allow us to determine whether the axisymmetry persists over a large range of particle sizes and whether resonant 
trapping of planetesimals dust by a migrating planet is indeed a viable explanation for the observed surface density of the disk
\citep{wya06}.

\section{Summary and Concluding Remarks\label{summary}}

Our ACS/HRC coronagraphic images of HD~92945 reveal an inclined axisymmetric debris disk comprising an inner ring $\sim 
2$\farcs0--3\farcs0 (43--65~AU) from the star and a faint outer disk with average F606W (Broad $V$) and F814W (Broad $I$) surface 
brightness profiles declining as $r^{-0.75}$ for $r = 3$\farcs0--5\farcs1 (65--110~AU) and $r^{-8}$ for $r = 5$\farcs1--6\farcs7 
(110--145~AU).   The sharp break in the profiles at 110~AU suggests that the disk is truncated at that distance.  The observed 
relative surface-density profile is peaked at both the inner ring and the outer edge of the disk.  This morphology is unusual among
the 15 other disks that have been spatially resolved in scattered light, which typically exhibit either solitary rings with sharp 
edges (e.g., HR~4796A and Fomalhaut) or asymmetric nebulosity with indefinite outer limits (e.g., $\beta$~Pic and HD~61005).  Only 
HD~181327 has an axisymmetric ring and extended nebulosity somewhat akin to HD~92945's disk, but its nebulosity is asymmetric and 
has a steep (but not truncated) surface brightness profile within $\sim 450$~AU \citep{sch06}.  

The dust in HD~92945's outer disk scatters neutrally and isotropically in the $V$ and $I$ bands.  These characteristics contradict 
current optical models of compact and porous grains, which predict that grains larger than a few microns are neutral scatterers 
and submicron-sized grains are isotropic scatterers at these wavelengths.  The disk's anomalously low $V$-band albedo ($\sim 10$\%) 
also suggests a large population of submicron-sized grains.   If grains smaller than a few microns are absent, then stellar
radiation pressure may be the cause only if the dust is composed of highly absorptive materials like graphite.  Optical models of 
compact silicate grains suggest a maximum blow-out size of $\sim 0.25$~\micron, and this size decreases as porosity increases 
\citep{muk92}.

Our {\it Spitzer} MIPS and IRS measurements reveal no significant infrared excess from HD~92945's disk shortward of 30~\micron, and 
they constrain the width of the 70~\micron\ source to $\lesssim 180$~AU.   Assuming that the dust comprises compact grains of 
astronomical silicate confined to the disk imaged with ACS, we modeled the 24--350~\micron\ emission with a grain-size distribution 
$dn/da \propto a^{-3.7}$ and $a_{min} = 4.5$~\micron\ throughout the disk, but with $a_{max} = 900$~\micron\ and 50~\micron\ in the inner 
ring and outer disk, respectively.  Combining these thermal constraints with the albedo and surface-density profile obtained from 
our ACS images, we obtain an integrated dust mass of $\sim 0.001~M_{\earth}$.

Conflicting indicators of minimum grain size are not unique to HD~92945.  \citet{kri10} reported that the narrow ring around the 
G0~{\small V} star HD~207129 exhibits both isotropic scattering and very low albedo ($\sim 5$\%), but its far-infrared SED is 
adequately modeled with silicate grains with $a_{min} = 2.8$~\micron\ and $dn/da \propto a^{-3.9}$.   As in the case 
of HD~92945, this value of $a_{min}$ greatly exceeds the radiative blow-out size for the host star, and $dn/da$ 
is significantly steeper than the traditional $a^{-3.5}$ representation for a steady-state collisional cascade \citep{doh69}.  The 
results for HD~92945 and HD~207129 are based on reliable and well-calibrated {\it HST} and {\it Spitzer} data, so it appears that 
the conflicting indicators stem from an incomplete understanding of the composition and optical properties of compact and porous 
grains around low-mass stars.  Current grain models indicate that the scattering asymmetry parameter and albedo are insufficiently 
sensitive to composition and porosity to account for our observational results, but some contradictory trends demand that caution 
be exercised when applying the models to observational data.  As demonstrated by \citet{gra07} for AU~Mic, polarimetric imaging of 
HD~92945's disk may provide definitive constraints on the sizes and porosity of the grains and thus avoid some ambiguities of the 
grain models.

HD~92945's disk morphology is remarkably like those predicted from dynamical models of dust produced from collisions of planetesimals
perturbed by coalescing or migrating planets.  The planet-resonance model of \citep{wya06} is particularly intriguing because it yields,
for a plausible range of grain sizes, a double-peaked surface density profile that resembles the observed profile of HD~92945's disk.  
Furthermore, this model predicts differences between the size distributions of grains that remain in the resonance in which they were 
created and those that leave the resonance on bound, axisymmetric orbits because of radiation pressure.   Such spatial segregation by 
grain size may be relevant to our thermal model of HD~92945's infrared SED, which requires two components of dust distinguished by 
values of $a_{max}$ that are 18 times larger in the inner ring than in the outer disk in order to match the observed 70, 160, and 
350~\micron\ fluxes.

As \citet{wya06} has advocated and as existing multiband images of the disks around $\beta$~Pic, Vega, and Fomalhaut have shown, 
resolved images over a broad spectral range are needed to constrain the composition and location of the HD~92945's dust, as well 
as the mechanism(s) responsible for the disk's morphology at each wavelength.   The high-resolution infrared and millimeter imaging 
capabilities of the {\it Herschel} Observatory and the Atacama Large Millimeter Array (ALMA) are well suited for determining the 
location or distribution of the unresolved thermal emission detected with {\it Spitzer}.  If the planet-resonance model of 
\citet{wya06} applies to HD~92945, then {\it Herschel} and ALMA images should reveal increasing concentrations of resonant dust 
as the imaging wavelength increases.   Constraining the location of the resonances would in turn constrain the mass and location 
of a putative migrating planet, which, at an age of $\sim 300$~Myr and possible mass of only a few $M_{\earth}$ (\S\ref{modcomp}), 
may not be directly detected with a high-resolution, near-infrared coronagraph such as that used to image the younger giant planet 
$\beta$~Pic~b \citep{lag10}.

Given the demise of the ACS/HRC and the uncertain future of NICMOS, the next likely opportunity for imaging HD~92945's disk in 
scattered light will follow the launch of the {\it James Webb Space Telescope.}  {\it JWST's} Near-Infrared Camera and Tunable 
Filter Imager will provide coronagraphic imaging from 1.5--5.0~\micron, which together with our ACS images will permit an assessment
of the chromatic dependence of the albedo, color, and scattering asymmetry of the dust over nearly a decade of wavelengths.   
Coronagraphic imaging with {\it JWST's} Mid-Infrared Instrument (MIRI) will probably be less fruitful, as {\it Spitzer} observations 
have shown no significant excess flux from dust shortward of 30~\micron.  However, MIRI will be useful for assessing the presence
and characteristics of an infant planetary system.   That said, more immediately accessible 450 and 850~\micron\ ground-based 
photometry would help to constrain the submillimeter end of HD~92945's SED and, consequently, the distribution and size limits of 
grains in the inner ring and outer disk.  \\

\acknowledgments 
We gratefully acknowledge Paul Smith from the University of Arizona for his assistance with MIPS SED data processing.  We also thank 
Glenn Schneider and collaborators for sharing the results of their NICMOS observations of HD~92945 prior to publication.  ACS was developed 
under NASA contract NAS~5-32865, and this research has been supported by NASA grant NAG5-7697 to the ACS Investigation Definition Team.   
Additional support for John Krist and Karl Stapelfeldt was provided by NASA through grants HST-GO-10539 and HST-GO-10854.  This research was 
partially supported by NASA through JPL/Caltech contract 1255094 to the University of Arizona.  It made use of Tiny Tim/Spitzer, developed 
by John Krist for the Spitzer Science Center, which is managed by Caltech under a contract with NASA.   The Space Telescope Science Institute
is operated by AURA Inc., under NASA contract NAS5-26555.

\newpage 

\begin{deluxetable}{llcccllc}
\tabletypesize{\footnotesize}
\tablecaption{Sixteen debris disks imaged in scattered light\label{16disks}}
\tablehead{
                    & \colhead{Spectral} & \multicolumn{2}{c}{Fractional Luminosity}        & \hspace*{0.1in} & \multicolumn{3}{c}{Imaging Information\tablenotemark{a}} \\
\cline{3-4}\cline{6-8}\\[0.001in]
\colhead{Host Star} & \colhead{Type}     & \colhead{log($L_{\rm IR}/L_*$)} & \colhead{Ref.} &                 & \colhead{Telescope} &\colhead{Coronagraph} & \colhead{Ref.} 
}
\tablecolumns{8}
\startdata 
$\beta$~Pic         & A5~V               & $-2.82$                         & 1              &                 & Dupont 2.5~m        & Seeing-limited       & 9 \\
HR~4796A            & A0~V               & $-2.3$                          & 2              &                 & {\it HST}           & NICMOS               & 10 \\
HD~141569A          & B9.5~Ve            & $-2.12$                         & 3              &                 & {\it HST}           & NICMOS               & 11,12 \\
AU~Mic	            & M1~Ve              & $-3.36$                         & 4              &                 & UH 2.2~m            & Seeing-limited       & 13 \\
HD 107146           & G2~V               & $-3.1$                          & 5              &                 & {\it HST}           & ACS/HRC              & 14 \\
Fomalhaut           & A3~V               & $-4.34$                         & 6              &                 & {\it HST}           & ACS/HRC              & 15 \\
HD~32297            & A0~V               & $-2.55$                         & 6              &                 & {\it HST}           & NICMOS               & 16 \\
HD~53143            & K1~V               & $-3.60$                         & 7              &                 & {\it HST}           & ACS/HRC              & 17 \\
HD~139664           & F5~V               & $-4.05$                         & 7              &                 & {\it HST}           & ACS/HRC              & 17 \\
HD~181327           & F5.5~V             & $-2.77$                         & 8              &                 & {\it HST}           & NICMOS, ACS/HRC      & 8 \\
HD~15115            & F2~V               & $-3.30$                         & 7              &                 & {\it HST}           & ACS/HRC              & 18 \\
HD~15745            & F2~V               & $-2.92$                         & 7              &                 & {\it HST}           & ACS/HRC              & 19 \\
HD~61005            & G8~V               & $-2.6$                          & 5              &                 & {\it HST}           & NICMOS               & 20 \\
HD~207129           & G0~V               & $-3.85$                         & 7              &                 & {\it HST}           & ACS/HRC              & 21 \\
HD~10647            & F9~V               & $-3.52$                         & 7              &                 & {\it HST}           & ACS/HRC              & 22 \\
HD~92945            & K1~V               & $-3.12$                         & 4              &                 & {\it HST}           & ACS/HRC              & 23 \\
\enddata

\tablenotetext{a}{ Pertains to the first published scattered-light images of the disks. \newline
REFERENCES.-- (1) \citealt{dec03}; (2) \citealt{jur91}; (3) \citealt{zuc95}; (4) \citealt{pla09}; (5) \citealt{hil08}; (6) \citealt{sil00}; (7) \citealt{zuc04}; (8) \citealt{sch06}; (9) \citealt{smi84}; (10) \citealt{sch99}; (11) \citealt{aug99}; (12) \citealt{wei99}; (13) \citealt{kal04}; (14) \citealt{ard04}; (15) \citealt{kal05a}; (16) \citealt{sch05}; (17) \citealt{kal06}; (18) \citealt{kal07b}; (19) \citealt{kal07a}; (20) \citealt{hin07}; (21) \citealt{kri10}; (22) \citealt{sta11}; (23) this paper. }

\end{deluxetable}

\clearpage

\clearpage

\clearpage

\clearpage

\clearpage

\clearpage

\clearpage

\clearpage

\clearpage

\clearpage


\begin{thebibliography}{}

\bibitem[Ardila et al.(2004)]{ard04} Ardila, D.~R., et al. 2004, \apj, 617, L147
\bibitem[Augereau et al.(1999)]{aug99} Augereau, J.~C., Lagrange, A.-M., Mouillet, D., \& M\'{e}nard, F. 1999, \aap, 350, L51
\bibitem[Backman \& Paresce(1993)]{bac93} Backman, D.~E., \& Paresce, F. 1993, in Protostars and Planets III, ed.\ E.~H.\ Levy \& J.~I.\ Lunine (Tucson: University of Arizona), 1253 
\bibitem[Burns et al.(1979)]{bur79} Burns, J.~A., Lamy, P.~L., \& Soter, S. 1979, Icarus, 40, 1
\bibitem[Campo Bagatin et al.(1994)]{cam94} Burns, J.~A., Lamy, P.~L., \& Soter, S. 1979, Icarus, 40, 1
\bibitem[Castelli \& Kurucz(2003)]{cas03} Castelli, F., \& Kurucz, R.~L. 2003, in Proc.\ IAU Symp. 210, eds.\ N.~Piskunov, W.~W.\ Weiss, \& D.~F.\ Gray (San Francisco: ASP), A20
\bibitem[Chiang et al.(2009)]{chi09} Chiang, E., Kite, E., Kalas, P., Graham, J.~R., \& Clampin, M. 2009, \apj, 693, 734
\bibitem[Chen et al.(2005)]{che05} Chen, C.~H., et al. 2005, \apj, 634, 1372
\bibitem[Clampin et al.(2003)]{cla03}  Clampin, M., et al. 2003, \aj, 126, 385
\bibitem[Colbert et al.(2010)]{col10} Colbert, J., et al. 2010, MIPS Instrument Handbook, Version 2.0 (Pasadena: SSC)
\bibitem[Decin et al.(2003)]{dec03} Decin, G., Dominik, C., Waters, L.~B.~F.~M., \& Waelkens, C. 2003, \apj, 598, 636
\bibitem[Dohnanyi(1969)]{doh69} Dohnanyi, J.~W. 1969, \jgr, 74, 2531
\bibitem[ESA(1997)]{esa97} ESA, 1997, The Hipparcos and Tycho Catalogues, ESA SP-1200
\bibitem[Fitzgerald et al.(2007)]{fit07} Fitzgerald, M.~P., Kalas, P.~G., Duch\^{e}ne, G., Pinte, C., \& Graham, J.~R. 2007, \apj, 670, 536
\bibitem[Ford et al.(2003)]{for03}Ford, H.~C., et al. 2003, Proc.\ SPIE, 4854, 81
\bibitem[Gilliland(2004)]{gil04} Gilliland, R.~L. 2004, Instrument Science Report ACS 2004-01 (Baltimore: STScI)
\bibitem[Golimowski et al.(2006)]{gol06} Golimowski, D.~A., et al. 2006, \aj, 131, 3109
%\bibitem[Golimowski et al.(2006)]{gol06} Golimowski, D.~A., Ardila, D.~R., Krist, J.~E., Clampin, M., Ford, H.~C., Illingworth, G.~D.,  Bartko, F., Benitez, N., Blakeslee, J.~P., Bouwens, R.~J., Bradley, L.~D., Broadhurst, T.~J., Brown, R.~A., Burrows, C.~J., Cheng, E.~S., Cross, N.~J.~G., Demarco, R., Feldman, P.~D., Franx, M., Goto, T., Gronwall, C., Hartig, G.~F., Holden, B.~P., Homeier, N.~L., Infante, L., Jee, M.~J., Kimble, R.~A., Lesser, M.~P., Martel, A.~R., Mei, S., Menanteau, F., Meurer, G.~R., Miley, G.~K., Motta, V., Postman, M., Rosati, P., Sirianni, M., Sparks, W.~B., Tran, H.~D., Tsvetanov, Z.~I., White, R.~L., Zheng, W., \& Zirm, A.~W. 2006, \aj, 131, 3109
\bibitem[Graham et al.(2007)]{gra07} Graham, J.~R., Kalas, P.~G., \& Matthews, B.~C. 2007, \apj, 654, 595
\bibitem[Grigorieva et al.(2007)]{gri07} Grigorieva, A., Artymowicz, P., \& Th\'{e}bault, P. 2007, \aap, 461, 537
\bibitem[Gordon et al.(2005)]{gor05} Gordon, K.~D., et al. 2005, \pasp, 117, 503
\bibitem[Hage \& Greenberg(1990)]{hag90} Hage, J.~I., \& Greenberg, J.~M. 1990, \apj, 361, 251
\bibitem[Heim et al.(1998)]{hei98} Heim, G.~B., et al. 1998, Proc. SPIE, 3356, 985
\bibitem[Henyey \& Greenstein(1941)]{hen41} Henyey, L.~G., \& Greenstein, J.~L. 1941, \apj, 93, 70
\bibitem[Hillenbrand et al.(2008)]{hil08} Hillenbrand, L.~A., et al. 2008, \apj, 677, 630
\bibitem[Hines et al.(2007)]{hin07} Hines, D.~C., et al. 2007, \apj, 671, L165
\bibitem[Holland et al.(1998)]{hol98} Holland, W.~S., et al. 1998, \nat, 392, 788
\bibitem[Houck et al.(2004)]{hou04}Houck, J.~R., et al. 2004, Proc.\ SPIE, 5487, 62
\bibitem[Ingleby et al.(2009)]{ing09} Ingleby, L., et al. 2009, \apj, 703, L137
\bibitem[Jura(1991)]{jur91} Jura, M. 1991, \apj, 383, L79
\bibitem[Kalas(2005)]{kal05b} Kalas, P. 2005, \apj, 635, L169
\bibitem[Kalas et al.(2007a)]{kal07a} Kalas, P., Duchene, G., Fitzgerald, M.~P., \& Graham, J.~R. 2007, \apj, 671, L161
\bibitem[Kalas et al.(2007b)]{kal07b} Kalas, P., Fitzgerald, M.~P., \& Graham, J.~R. 2007, \apj, 661, L85
\bibitem[Kalas et al.(2005)]{kal05a} Kalas, P., Graham, J.~R., \& Clampin, M. 2005, \nat, 435, 1067
\bibitem[Kalas et al.(2006)]{kal06} Kalas, P., Graham, J.~R., Clampin, M., \& Fitzgerald, M.~P. 2006, \apj, 637, L57
\bibitem[Kalas et al.(2004)]{kal04} Kalas, P., Liu, M.~C., \& Matthews, B.~C. 2004, Science, 303, 1990
\bibitem[Kennedy \& Kenyon(2008)]{ken08} Kennedy, G.~M., \& Kenyon, S.~J. 2008, \apj, 673, 502
\bibitem[Kenyon \& Bromley(2002)]{ken02} Kenyon, S.~J., \& Bromley, B.~C. 2002, \apj, 577, L35
\bibitem[Kenyon \& Bromley(2004)]{ken04} Kenyon, S.~J., \& Bromley, B.~C. 2004, \aj, 127, 513
\bibitem[Klahr \& Lin(2000)]{kla00} Klahr, H.~H., \& Lin, D.~N.~C. 2000, \apj, 554, 1095
\bibitem[K\"{o}hler et al.(2007)]{koh07} K\"{o}hler, M., Minato, T., Kimura, H., \& Mann, I. 2007, Adv.\ Space Res., 40, 266
\bibitem[Krist(2000)]{kri00} Krist, J. 2000, Instrument Science Report ACS 2000-04 (Baltimore: STScI)
\bibitem[Krist(2002)]{kri02} Krist, J. 2002, Instrument Science Report ACS 2002-11 (Baltimore: STScI)
\bibitem[Krist(2006)]{kri06} Krist, J. 2006, Tiny Tim/Spitzer User's Guide (Pasadena: SSC)
\bibitem[Krist \& Hook(2004)]{kri04} Krist, J., \& Hook, R. 2004, The Tiny Tim User's Guide, Version 6.3 (Baltimore: STScI)
\bibitem[Krist et al.(2005)]{kri05} Krist, J.~E., et al. 2005, \aj, 129, 1008
\bibitem[Krist et al.(2010)]{kri10} Krist, J.~E., et al. 2010, \aj, 140, 1051
\bibitem[Krist et al.(2003)]{kri03} Krist, J.~E., Hartig, G.~F., Clampin, M., Golimowski, D.~A., Ford, H.~C., \& Illingworth, G.~D. 2003, Proc.\ SPIE, 4860, 20
\bibitem[Krivov et al.(2006)]{krv06} Krivov, A.~V., L\"{o}hne, T., \& Srem\v{c}evi\'{c}, M. 2006, \aap, 455, 509
\bibitem[Lagrange et al.(2000)]{lag00} Lagrange, A.-M., Backman, D.~E., \& Artymowicz, P. 2000, in Protostars and Planets IV, ed.\ V.\ Mannings, A.P.\ Boss, \& S.~S.\ Russell (Tucson: University of Arizona), 639 
\bibitem[Lagrange et al.(2010)]{lag10} Lagrange, A.-M., et al. 2010, Science, 329, 57 
\bibitem[Laidler et al.(2005)]{lai05} Laidler, V. et al. 2005, Synphot User's Guide, Version 5.0 (Baltimore: STScI)
\bibitem[Laor \& Draine(1993)]{lao93} Laor, A., \& Draine, B.~T. 1993, \apj, 402, 441
\bibitem[Li \& Greenberg(1998)]{li98} Li, A., \& Greenberg, J.~M. 1998, \aap, 331, 291
\bibitem[Li \& Lunine(2003a)]{li03a} Li, A., \& Lunine, J.~I. 2003a, \apj, 590, 368
\bibitem[Li \& Lunine(2003b)]{li03b} Li, A., \& Lunine, J.~I. 2003b, \apj, 594, 987
\bibitem[Li et al.(2003)]{llb03} Li, A., Lunine, J.~I., \& Bendo, G.~J. 2003, \apj, 598, L51
\bibitem[L\'{o}pez-Santiago et al.(2006)]{lop06} L\'{o}pez-Santiago, J., Montes, D., Crespo-Chac\'{o}n, I., \& Fern\'{a}ndez-Figueroa, M.~J. 2006, \apj, 643, 1160
\bibitem[Mamajek \& Hillenbrand(2008)]{mam08} Mamajek, E.~E., \& Hillenbrand, L.~A. 2008, \apj, 687, 1264
\bibitem[Maness et al.(2009)]{man09} Maness, H.~L., et al. 2009, \apj, 707, 1098
\bibitem[Maness et al.(2008)]{man08} Maness, H.~L., Fitzgerald, M.~P., Paladini, R., Kalas, P., Duchene, G.\& Graham, J.~R. 2008, \apj, 686, L25
\bibitem[Mawet et al.(2009)]{maw09} Mawet, D., Serabyn, E., Stapelfeldt, K., \& Crepp, J. 2009, \apj, 702, L47
\bibitem[Maybhate et al.(2010)]{may10} Maybhate, A., et al. 2010, ACS Instrument Handbook, Version 10.0 (Baltimore: STScI)
%\bibitem[Metchev et al.(2005)]{met05} Metchev, S.~A., Eisner, J.~A., Hillenbrand, L.~A., \& Wolf, S. 2005, \apj, 622, 451
\bibitem[Meurer et al.(2002)]{meu02} Meurer, G.~R., et al. 2002, in 2002 {\it HST} Calibration Workshop, eds.\ S.~Arribas, A.~Koekemoer, and B. Whitmore (Baltimore: STScI), 65
\bibitem[Meyer et al.(2007)]{mey07} Meyer, M.~R., Backman, D.~E., Weinberger, A.~J., \& Wyatt, M.~C. 2007, in Protostars and Planets V, eds.\ B.~Reipurth, D.~Jewitt, and K.~Keil (Tucson: University of Arizona), 573 
\bibitem[Mukai et al.(1992)]{muk92} Mukai, T., Ishimoto, H., Kozasa, T., Blum, J., \& Greenberg, J.~M. 1992, \aap, 262, 315
\bibitem[M\"{u}ller et al.(2010)]{mul10} M\"{u}ller, S., L\"{o}hne, T., \& Krivov, A.~V. 2010, \apj, 708, 1728
\bibitem[Pavlovsky et al.(2006)]{pav06} Pavlovsky, C., et al. 2006, ACS Data Handbook, Version 5.0 (Baltimore: STScI)
\bibitem[Plavchan et al.(2005)]{pla05} Plavchan, P., Jura, M., \& Lipscy, S.~J. 2005, \apj, 631, 1161
\bibitem[Plavchan et al.(2009)]{pla09} Plavchan, P., Werner, M.~W., Chen, C.~H., Stapelfeldt, K.~R., Su, K.~Y.~L., Stauffer, J.~R., \& Song, I. 2009, \apj, 698, 1068
\bibitem[Saija et al.(2003)]{sai03} Saija, R., Iat\`{i}, M.~A., Giusto, A., Borghese, F., Denti, P., Aiello, S., \& Cecchi--Pestellini, C. 2003, \mnras, 341, 1239
\bibitem[Schneider et al.(1999)]{sch99} Schneider, G., et al. 1999, \apj, 513, L127
\bibitem[Schneider et al.(2006)]{sch06} Schneider, G., et al. 2006, \apj, 650, 414
\bibitem[Schneider et al.(2005)]{sch05} Schneider, G., Silverstone, M.~D., \& Hines, D.~C. 2005, \apj, 629, L117
\bibitem[Shen et al.(2009)]{she09} Shen, Y., Draine, B.~T., \& Johnson, E.~T. 2009, \apj, 696, 2126
\bibitem[Silverstone(2000)]{sil00} Silverstone, M.~D. 2000, Ph.D. thesis, Univ.\ California at Los Angeles
\bibitem[Sirianni et al.(2005)]{sir05} Sirianni, M., et al. 2005, \pasp, 117, 1049
\bibitem[Skrutskie et al.(1997)]{skr97} Skrutskie, M.~F., et al.\ 1997, in The Impact of Large Scale Near-IR Surveys, ed.\ F.~Garzon (Dordrecht: Kluwer), 25
\bibitem[Smith et al.(1992)]{smi92} Smith, B.~A., Fountain, J.~W., \& Terrile, R.~J. 1992, A\&A, 261, 499
\bibitem[Smith \& Terrile(1984)]{smi84} Smith, B.~A., \& Terrile, R.~J. 1984, Science, 226, 1421
\bibitem[Stapelfeldt et al.(2011)]{sta11} Stapelfeldt, K.~R., et al. 2011, in prep.
\bibitem[Strubbe \& Chiang(2006)]{str06} Strubbe, L., \& Chiang, E. 2006, \apj, 648, 652
\bibitem[Takeuchi \& Artymowicz(2001)]{tak01} Takeuchi, T., \& Artymowicz, P. 2001, \apj, 557, 990
\bibitem[Thatte et al.(2009)]{tha09} Thatte, D., et al. 2009, NICMOS Data Handbook, Version 8.0 (Baltimore: STScI)
\bibitem[Th\'{e}bault \& Augereau(2007)]{the07} Th\'{e}bault, P., \& Augereau, J.-C. 2007, \aap, 472, 169
\bibitem[Th\'{e}bault et al.(2003)]{the03} Th\'{e}bault, P., Augereau, J.-C., \& Beust, H. 2003, \aap, 408, 775
\bibitem[Th\'{e}bault \& Wu(2008)]{the08} Th\'{e}bault, P., \& Wu, Y. 2008, \aap, 481, 713
\bibitem[Viana et al.(2009)]{via09} Viana, A., et al. 2009, NICMOS Instrument Handbook, Version 11.0 (Baltimore: STScI)
\bibitem[Voshchinnikov et al.(2005)]{vos05} Voshchinnikov, N.~V., Il'in, V.~B., \& Henning, T. 2005, \aap, 429, 371
\bibitem[Weinberger et al.(1999)]{wei99} Weinberger, A.~J., Becklin, E.~E., Schneider, G., Smith, B.~A., Lowrance, P.~J., Silverstone, M.~D., Zuckerman, B., \& Terrile, R.~J. 1999, \apj, 525 L53
\bibitem[Wood et al.(2005)]{woo05} Wood, B.~E., M\"{u}ller, H.-R., Zank, G.~P., Linsky, J.~L., \& Redfield, S. 2005, \apj, 628, L143
\bibitem[Wyatt(2003)]{wya03}Wyatt, M.~C. 2003, \apj, 598, 1321
\bibitem[Wyatt(2006)]{wya06}Wyatt, M.~C. 2006, \apj, 639, 1153
\bibitem[Wyatt(2008)]{wya08}Wyatt, M.~C. 2008, \araa, 46, 339
\bibitem[Zuckerman(2001)]{zuc01} Zuckerman, B. 2001, \araa, 39, 549
\bibitem[Zuckerman et al.(1995)]{zuc95} Zuckerman, B., Forveille, T., \& Kastner, J.~H. 1995, \nat, 373, 494
\bibitem[Zuckerman \& Song(2004)]{zuc04} Zuckerman, B., \& Song, I. 2004, \apj, 603, 738

\end{thebibliography}
\end{document}